\documentclass[12pt]{iopart}
\usepackage{graphicx}
\usepackage{bm}
\usepackage{color}
\begin{document}

\title{Nonlinear acousto-magneto-plasmonics}

\author{Vasily V. Temnov}
\address{IMMM CNRS 6283, Universit\'e du Maine, 72085 Le Mans
cedex, France}
\address{Fritz-Haber-Institut der MPG, Phys. Chemie,
Faradayweg 4-6, 14195 Berlin, Germany}
\author{Ilya Razdolski}
\address{Fritz-Haber-Institut der MPG, Phys. Chemie,
Faradayweg 4-6, 14195 Berlin, Germany}
\author{Thomas Pezeril}
\address{IMMM CNRS 6283, Universit\'e du Maine, 72085 Le Mans
cedex, France}
\author{Denys Makarov}
\address{Helmholtz-Zentrum Dresden-Rossendorf e. V., Institute
of Ion Beam Physics and Materials Research, 01328 Dresden,
Germany}
\author{Denis Seletskiy}
\address{Department of Physics and Center for Applied
Photonics, University of Konstanz, D-78457 Konstanz, Germany}
\author{Alexey Melnikov}
\address{Fritz-Haber-Institut der MPG, Phys. Chemie,
Faradayweg 4-6, 14195 Berlin, Germany}
\author{Keith A. Nelson}
\address{Department of Chemistry, Massachusetts Institute
of Technology, Cambridge, Massachusetts 02139, USA}

\ead{vasily.temnov@univ-lemans.fr} \vspace{10pt}
\begin{indented}
\item[]February 2016
\end{indented}

\begin{abstract}
We review the recent progress in experimental and theoretical
research of interactions between the acoustic, magnetic and
plasmonic transients in hybrid metal-ferromagnet multilayer
structures excited by ultrashort laser pulses. The main focus is
on understanding the nonlinear aspects of the acoustic dynamics in
materials as well as the peculiarities in the nonlinear optical
and magneto-optical response. For example, the nonlinear optical
detection is illustrated in details by probing the static
magneto-optical second harmonic generation in gold-cobalt-silver
trilayer structures in Kretschmann geometry. Furthermore, we show
experimentally how the nonlinear reshaping of giant ultrashort
acoustic pulses propagating in gold can be quantified by
time-resolved plasmonic interferometry and how these ultrashort
optical pulses dynamically modulate the optical nonlinearities.
The effective medium approximation for the optical properties of
hybrid multilayers facilitates the understanding of novel optical
detection techniques. In the discussion we highlight recent works
on the nonlinear magneto-elastic interactions, and strain-induced
effects in semiconductor quantum dots.
\end{abstract}

%
%
%
\maketitle
%
\ioptwocol

\section{Motivation}

The main objective of this review is to summarize recent
developments in the field of the nonlinear interactions in the
magneto-plasmonic and acousto-plasmonic multilayer structures. We
are aiming to provide understanding of these fundamental
interactions governing the macroscopic properties of the complex
multilayer structures excited by femtosecond laser pulses. The
role of the ultrashort laser pulses is twofold. On the one hand,
high peak intensity is advantageous for efficient excitation of
the nonlinear optical processes such as second and third harmonic
generation. On the other hand, short laser pulses provide a
possibility for an excitation of the electronic, acoustic and
magnetic transients as well as allow for sampling of the ensuing
dynamics in pump-probe experiments. In addition to application of
the external magnetic field, which provides flexible control of
the static and dynamic properties in condensed matter systems, in
this review we emphasize the high potential that the nanoscale
acoustic perturbations offer for study as well as selective
dynamic control of various degrees of freedom in complex matter.
As we outline below, the functional nanolayers are particularly
attractive for these tasks as they support excitation and
propagation conditions for ultrashort acoustic pulses while at the
same time offer strong sub-micrometer confinement of
surface-plasmon-polaritons (SPPs). In turn, macroscopic
propagation of SPPs in nanolayers for distances exceeding tens of
micrometers makes it possible to enhance the non-linear
light-matter interactions in such systems.

\begin{figure}
\includegraphics [width= 8cm]{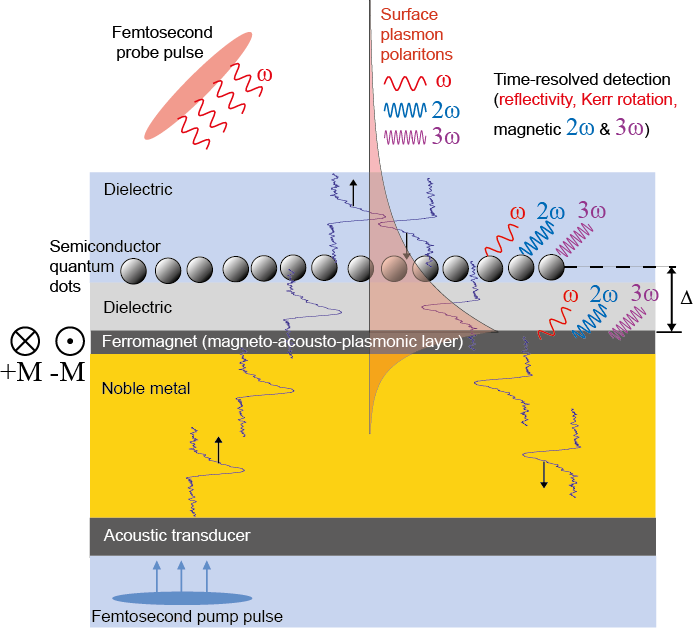}
\caption{Schematic of a prototypical device for studies of
magneto-acousto-plasmonic response of various material systems
(e.g. semiconductor quantum dots) under femtosecond excitation and
external magnetic field. Ultrafast pump pulse excites a structure
made from acoustic transducer, noble metal and a ferromagnet and
the ensuing dynamics is probed via various detection methods
reviewed here.}
\end{figure}

A prototypical experimental setup which embodies the concepts
outlined above is schematically depicted in Figure 1. Here, a
femtosecond pump pulse excites the nanolayer structure, which is
comprised of an acoustic transducer, noble metal and ferromagnetic
layer, while under an optional static magnetic field. Launched
magnetic, acoustic and plasmonic transient(s) can be used to study
ultrafast dynamics in a system of interest, which is embedded into
a dielectric cap layer in close proximity (at a distance $\Delta$)
to the metal-dielectric interface
\cite{Fedutik07PRL99,Ueda08APL92}. The two examples that we
motivate here is the control of the magneto-optical properties in
a ferromagnetic compound and time-domain studies of semiconductor
quantum dots under strong transient acoustic strain bias at the
nanoscale. The response of these systems can be probed via the
reviewed detection methods, such as magnetic second harmonic
generation (mSHG), time-resolved plasmonic and nonlinear optical
(second and third harmonic) probing.

While the device presented in Fig.~1 has not yet been realized, we
hope to convince the reader of its great potential for scientific
investigations of magneto-acousto-plasmonic interactions in
various nanoscale systems. As such, this review provides a
progress report on the recent studies which address various
isolated components or ideas, comprising potential building blocks
toward our envisioned device.

The presented material is divided into several sections. Section
two discusses two experimental geometries used to excite surface
plasmon polaritons in magneto-plasmonic multilayers: Kretschmann
configuration for a thin multilayer on a dielectric prism and the
plasmonic interferometry for macroscopically thick all-metallic
samples.

Section three describes some recent advances in the linear and
nonlinear magneto-plasmonics. A weak external magnetic field is
shown to induce large modulation in the intensity of surface
second harmonic generation (SHG) and also to identify the role of
surface plasmon polaritons in the nonlinear frequency conversion
phenomena. The effective medium approximation for
magneto-plasmonic metal-ferromagnet multilayer structures helps to
understand the results. A phenomenological model based on the
interference between the magnetic and nonmagnetic contributions
from two effective interfaces is developed to explain the
observations.

Section four discusses generation and characterization of giant
and ultrashort acoustic pulses in (noble metal)-ferromagnet
bilayer structures and evidences their impact on the SHG output.
Another phenomenological model based on the acoustic modulation of
the optical properties of a thin ferromagnetic layer is proposed
to explain the experimental observations.

Section five concerns the specificity of the sample fabrication
for plasmonic and acoustic measurements.

In the discussion section we outline other involved phenomena,
notably related to ultrafast magnetization dynamics, thermal
effects and strain effects in semiconductor quantum dots.

\section{Optical spectroscopy with surface plasmon polaritons}

Time-dependent perturbations of electron density distribution in
metals are at the heart of plasmonics. The relaxation dynamics is
governed by oscillations of the electrons at the so-called plasma
frequency, which, in the long-wavelength limit of free-electron
approximation, is given by $\omega_{\rm
p}=\sqrt\frac{n_ee^2}{\varepsilon_0m}$, where $n_e$, $e$ and $m$
are the electronic density, charge and mass, respectively and
$\varepsilon_0$ is the dielectric permittivity of free space.
Since in most metals there is at least one free electron per ion,
the resulting electron plasma density is very high, $n_e\sim
10^{22}$~cm$^{-3}$, with a corresponding $\omega_{\rm p}$ in the
visible to ultraviolet frequency range. Optical fields at
frequency $\omega < \omega_{\rm p}$ are effectively screened by
the plasma, resulting in a penetration profile which is
exponentially attenuated within the so-called {\it skin depth}
$\delta_{\rm skin}$. For the majority of metals the skin depth
lies in the range of 10-20 nm in the visible and near-infrared
frequency range. A Drude model provides the dielectric response
function of free electrons $\varepsilon_m(\omega)=1-\omega_{\rm
p}^2/(\omega(\omega+i/\tau_c))$, where $\tau_c$ stands for an
effective electron scattering time. This simple expression offers
a reasonable approximation for the optical properties of
free-carrier-like metals (Al,~Ag,~Au,~Cu) used in plasmonics
because of their small losses (large $\tau_c$).

In low-dimensional metallic nanostructures such as metal surfaces
(2D), wires (1D) or dots (0D), the frequency of plasma
oscillations is substantially affected by charge separation at the
surface. This is illustrated in Fig.~2 which shows the different
plasma frequencies and mixed electromagnetic/surface charge
density excitations in various geometries.

 \begin{figure}
   \includegraphics[width=8cm]{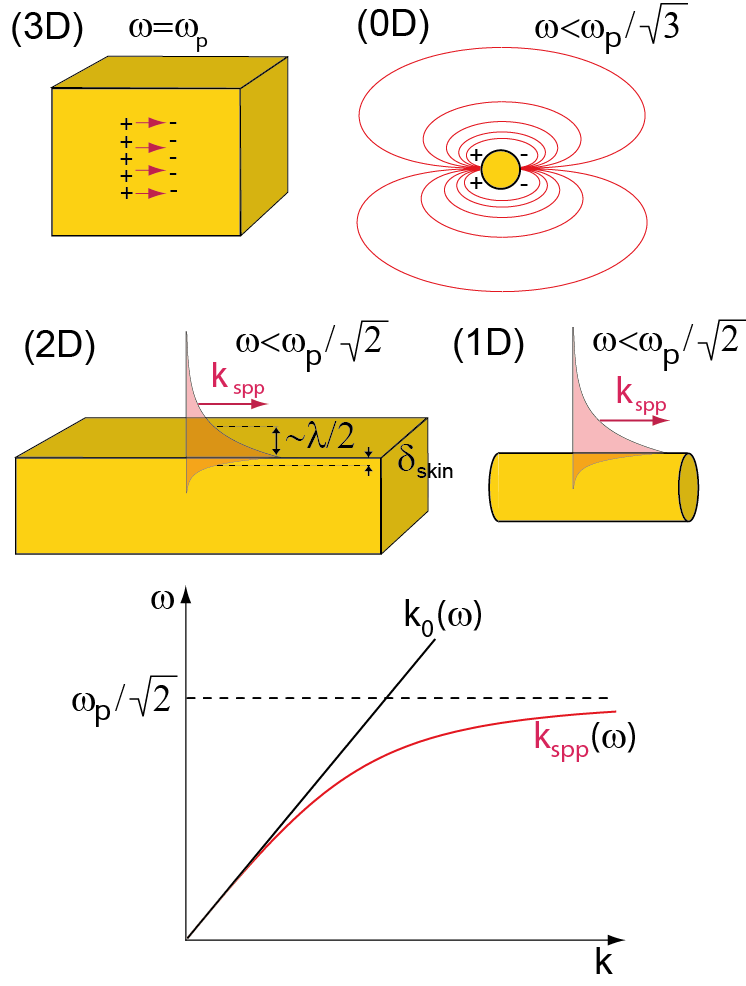}
    \caption[Surface plasmons in OD, 1D, 2D and 3D]
      {Frequency of oscillations of electron gas in bulk metal (3D), is modified by surface effects in metallic nanoparticles (0D), nanowires (1D) and surfaces (2D). Whereas metallic nanoparticles possess a specific, localized surface plasmon (LSP) resonance frequency determined by their geometry, surface plasmon polaritons (SPPs) in 1D and 2D form a broad frequency band $\omega(k_{\rm spp})$ and may propagate over macroscopic distances while obeying a characteristic dispersion relation.}
    \label{Dimensions}
  \end{figure}

In the case of zero-dimensionality (0D), the electron gas
oscillates at a fixed resonance frequency and may lead to
significant enhancement of optical fields close to the surface of
the nanoparticle due to the nanoantenna effect. This field
enhancement is of key importance to understanding the interaction
mechanisms between plasmonic nanoparticles and nano-scale light
emitters. The resonance frequency for metal nanoparticles of the
diameter smaller than the skin depth $\delta_{\rm skin}$ tends to
a value of $\omega_{\rm p}/\sqrt{3}$, while with the increased
diameter it experiences a red shift \cite{Pelton08LPhotRev2}. By
varying the diameter of 0D metallic nanoparticles ranging from a
few to tens of nanometers, this localized plasmon frequency can be
tuned over the entire visible spectral range.

In addition to the bulk and the localized plasmon modes, surface
plasmon modes can be supported at one or two dimensional
interfaces. These collective excitations are often referred to as
surface plasmon polaritons. These are coupled electromagnetic -
surface charge density waves on flat metal surfaces and wires
which can propagate over macroscopic distances, largely exceeding
the optical wavelength $\lambda$. In the case of metallic surface,
the electromagnetic fields in the direction normal to the
interface are exponentially bound. On the vacuum side, the typical
exponential decay length is a fraction of $\lambda$, therefore
resulting in strong sub-wavelength field confinement. Inside the
metal, the intensity of electric and magnetic components of the
electromagnetic field decays exponentially within the skin depth.
SPPs exist in a broad frequency band below the cut-off frequency
$\omega_{\rm p}/\sqrt{2}$ and are characterized by dispersion
relation $\omega(k_{\rm spp})$, where $k_{\rm spp}$ is the wave
vector of the surface plasmon (lower panel in Fig.~2). This
dispersion relation is identical for 1D and 2D cases, as long as
the nanowire radius significantly exceeds the skin depth
\cite{Wang06PRL96} and is given by $k_{\rm
spp}(\omega)=k_0(\omega)\sqrt\frac{\varepsilon_m(\omega)}{\varepsilon_m(\omega)+1}$,
where $k_0(\omega)=\omega/c$ is the wave vector of light in
vacuum.

\begin{figure}
    \includegraphics[width= 8cm]{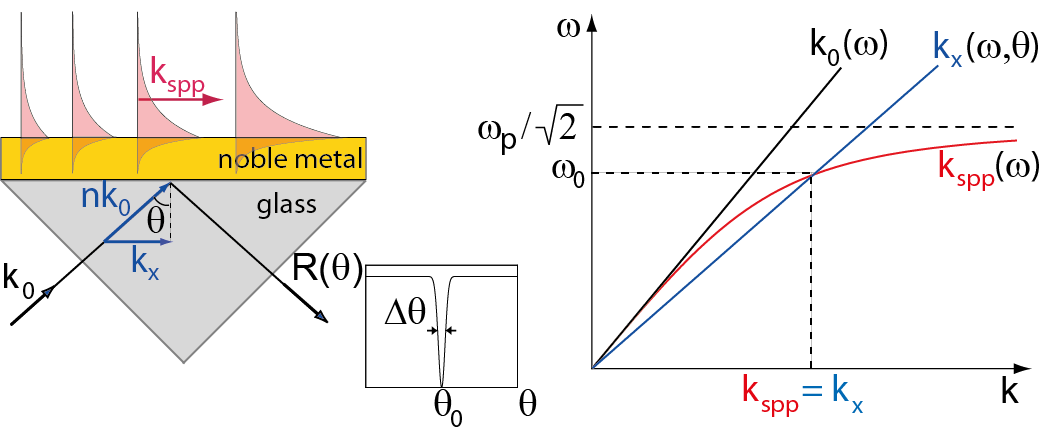}
    \caption[Excitation of surface plasmons in Kretschmann geometry]
      {Illumination of a thin (noble) metal film through the glass prism (known as Kretschmann geometry \cite{Kretschmann68})
      supports the conditions for excitation of the surface plasmon polariton modes. SPP excitation at the metal-air interface is manifested by a sharp minimum in the reflectivity $R(\theta)$, once the light is made incident at specific angle of $\theta_0$ of the SPP excitation condition. The condition is satisfied at the intersection point of the dispersion relations of the light and surface plasmon modes ($k_{\rm spp}(\omega_0)=k_x(\omega_0,\theta_0)$ on the right panel), resulting in strong electric field enhancement at the surface.}
    \label{Kretschmann}
  \end{figure}

The details of the coupling of the surface plasmon modes and
optical fields in low-dimensional structures are worth further
consideration. Due to the localized nature of the 0D resonance,
plasmonic response of the metallic nanoparticles can be
effectively approximated as a Hertzian dipole, where large spatial
bandwidth of the possible wave vectors manifests in direct
coupling plane electromagnetic waves. The situation is entirely
different for surface plasmon polaritons in 1D and 2D geometries.
Since their dispersion curve $\omega(k_{\rm spp})$ never crosses
the light line in vacuum ($k_{\rm spp}>k_0$ for all frequencies),
they cannot be excited by plane electromagnetic waves impinging on
a metal-air interface. An elegant and most widely used way to
excite SPP modes is provided by the so-called Kretschmann geometry
(see Fig.~3), where a thin metal film deposited on top of a
dielectric prism is illuminated by a collimated light beam through
the prism with refractive index $n>1$. For a particular
combination of the incident angle $\theta_0$ and light frequency
$\omega_0$, the in-plane component of the wave vector of incident
light $k_x(\omega,\theta)=n(\omega)k_0(\omega)\sin\theta$ matches
the wave vector of a surface plasmon polariton:
\begin{equation}
\label{LinearSPRcondition} k_{\rm
spp}(\omega_0)=n(\omega_0)k_0(\omega_0)\sin\theta_0\,.
\end{equation}

In case of a monochromatic laser source, the angular dependence of
reflectivity $R(\theta)$ shows a sharp dip at $\theta=\theta_0$
indicating that a significant fraction of incident energy is
converted into the SPP mode, propagating along the metal-air
interface. The metal layer should be relatively thin (typically
$\sim$50~nm) in order to allow for efficient coupling of the
incident light from the glass-metal to the the metal-air
interface. Thicker films would decrease the coupling efficiency
due to enhanced attenuation of the evanescent wave in metal
whereas in even thinner films the hybridization of the SPPs at the
two interfaces and the subsequent formation of symmetric and
antisymmetric modes would introduce additional radiative losses
\cite{BurkePRB1986}. Upon satisfaction of the SPP excitation
condition, the intensity of the excited SPP mode grows with
propagation along the illuminated region of the surface and can
get up to 300 times larger as compared to the incident wave,
limited by the ohmic losses in the metal film
\cite{Weber81OL6,Khurgin15}. In practice, this field enhancement
is lower due to parasitic effects which reduce the surface plasmon
propagation length such as, for example, surface roughness and
material inhomogeneities \cite{Nagpal09Science325}. In hybrid
ferromagnet-(noble metal) multilayer structures the field
enhancement is further reduced due to SPP absorption in a
ferromagnet.

\begin{figure}
    \includegraphics[width= 7cm]{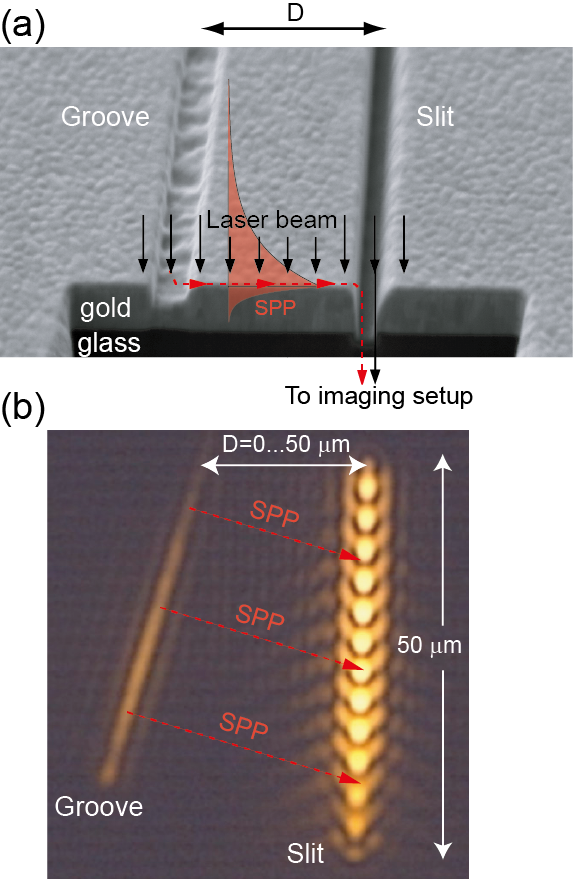}
    \caption[Excitation of surface plasmons in Kretschmann geometry]
      {Plasmonic microinterferometer based on a tilted slit-groove
pair. (a) SEM micrograph showing a typical depth profile of the
slit-groove structure (slit-groove distance $D$ of 1~$\mu$m,
slit-groove tilt angle of $3^{\circ}$, slit width 100~nm, groove
width 200~nm, groove depth 100~nm) milled into a 200~nm-thick gold
film on a glass substrate by a focused ion beam. (b) False-color
image of optical transmission of a homogeneously illuminated
microinterferometer in a near-infrared, with a minimum slit-groove
spacing of $D=20~\mu$m and slit-groove tilt angle of $15^{\circ}$.
The periodic modulation of light intensity transmitted through the
slit is due to interference with surface plasmons launched by the
groove (see text for details). Adopted with permission from
Ref.~\cite{SPInterferometry09OE}.}
    \label{SPInterferometry}
  \end{figure}

Scattering of the incident light on natural or artificial surface
defects, with a size of the order of a wavelength or smaller,
represents another common way to satisfy the condition for SPP
excitation \cite{ZayatsPhysRep05}. A pair of appropriately-spaced
sub-wavelength defects can serve for in- and outcoupling of the
incident light to the SPP modes and producing the plasmonic
interference fringes
\cite{Gay06NPh2_262,Temnov07OL32,Pacifici08NPhot1}.

Figure 4(a) shows a scanning electron microscope (SEM) micrograph
of the plasmonic microinterferometer, which consists of a
slit-groove pair milled into a 200~nm gold film with a 30~kV
Ga$^+$ focused ion beam. The 200~nm gold film was grown by
magnetron sputtering on a 2~nm-thin chromium buffer layer on a
glass substrate. The length of both the slit and the groove is
50~$\mu$m. The slit has a width of 100~nm and extends through the
thickness of the gold layer to the glass substrate. The depth of
the groove is 100~nm and it is 200~nm wide. A series of similar
structures was fabricated, where the slit-groove tilt angle was
systematically varied between 3$^{\circ}$ and 15$^{\circ}$  and
the slit-groove distance between 0 and 50~$\mu$m. If the whole
area of the microinterferometer is homogeneously illuminated with
coherent light, the SPP, first launched at the groove, will
propagate toward the slit and interfere with the directly
transmitted light upon outcoupling back into the light mode. An
optical transmission image of a slit-groove microinterferometer,
illuminated by a collimated p-polarized laser beam (beam waist
$\sim40~\mu$m FWHM, $\lambda=800~$nm) under nearly normal
incidence, is shown in Fig. 4(b). A bright periodic intensity
modulation of light transmitted through the slit represents the
plasmonic interference pattern. A small tilt angle between the
slit and the groove determines the spatial periodicity of
interference fringes and allows for interferometric measurements
at a single wavelength with a single microinterferometer.

The interferometric sensitivity to the phase shift acquired by the
SPP upon macroscopic propagation of the distance between the
groove and the slit can be harnessed for sensing. In the hybrid
multilayers, the phase and/or the contrast of these plasmonic
interference fringes can be conveniently controlled via the
modulation of the dielectric susceptibility by means of either
magneto-optical effects in magneto-plasmonics or elasto-optical
effects in acousto-plasmonics.

\section{Magneto-plasmonics}
\subsection{Linear magneto-plasmonics}

The influence of the external magnetic field on plasmonic
properties is extensively discussed elsewhere
\cite{ArmellesAdvOptMater13}. The magnetic field control of the
plasmonic properties has been demonstrated in a number of systems
\cite{CtistisNano09,DemidenkoJOSAB2011,BonanniNano11,
BelotelovNatCommun13,CrasseeNano12,MaccaferriPRL13,ShcherbakovPRB14,AutoreACS15,MaccaferriNatCommun15,MaksimovNanomat15}.
Here we shall focus on the properties of the SPPs in a hybrid
(noble) metal-ferromagnet-(noble) metal multilayer structure,
which serves as a robust playground for magneto-plasmonics
(Fig.~5). The geometry of SPPs at 800~nm wavelength in a Au-Co-Au
multilayer structure is shown in Fig.~5(a). Both the real and
imaginary parts of the complex SPP wave vector can be calculated
within the effective medium approximation
\cite{Temnov10NPhoton4,RazdolskiACS16}. The effective dielectric
function \begin{equation} \label{e_effective} \varepsilon_{\rm
eff}=\frac{1}{\delta_{\rm skin}}\int_0^{\infty}\varepsilon(z){\rm
e}^{-z/\delta_{\rm skin}}dz\,,
\end{equation}
accurately reproduces the SPP dispersion relation in a
magneto-plasmonic multilayer structure
\begin{equation}
\label{SP dispersion} k_{\rm spp}= k_0\sqrt{\frac{\varepsilon_{\rm
eff}}{1+\varepsilon_{\rm eff}}}\,.
\end{equation}

\begin{figure}
    \includegraphics[width= 8cm]{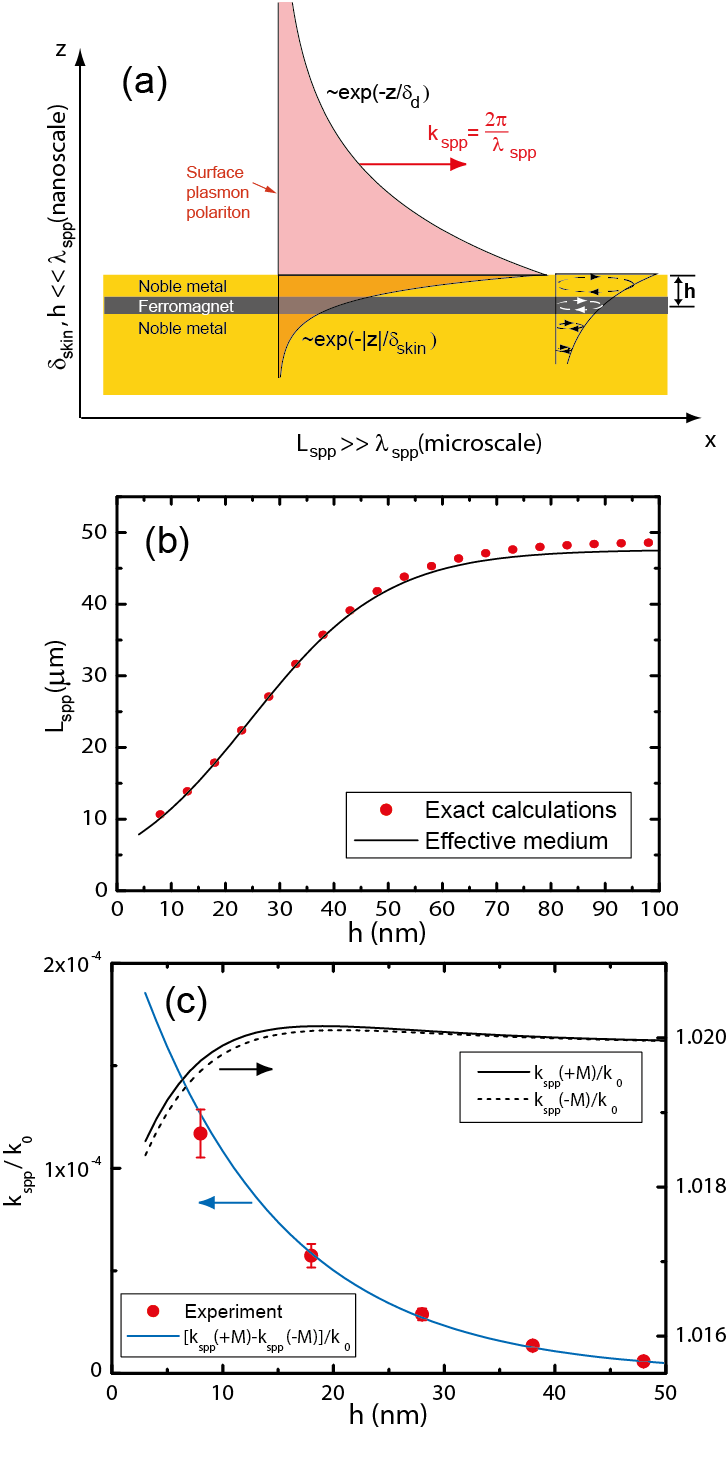}
    \caption[SPPs in hybrid metal-ferromagnet multilayer structures]
      {(a) SPPs in metal/ferromagnet/metal multilayer
structures possess the spatial distribution of the in-plane
electric field (shaded area shows $|E_z(z)|^2$), which is almost
identical to that for a single metal film. Dashed elliptical
contours represent the shape of the SPP electric field driving the
motion of the electrons.
(b) The dependence of the SPP decay length $L_{\rm spp}$ on the
location $h$ of the cobalt layer beneath the gold-air interface,
showing excellent agreement with an effective medium approximation
fit (black line). (c) In-plane magnetization reversal (in
$y$-direction) in ferromagnetic cobalt changes the SPP wave vector
$k_{\rm spp}(\pm M)$. Its magnetic modulation $k_{\rm
spp}(+M)-k_{\rm spp}(+M)$ obtained within the effective medium
approximation is in agreement with experimental measurements from
Ref.~\cite{Temnov10NPhoton4}. Figure (a) adopted with permission
from Ref.~\cite{Temnov12NPhoton6}.} \label{MP_multilayers}
\end{figure}

The validity of this approximation is illustrated by the
dependence of SPP propagation length $L_{\rm spp}=1/(2{\rm
Im}[k_{\rm spp}])$ in a Au-Co-Au multilayer, where the 6~nm-thick
cobalt layer was placed at a variable depth $h$ below the gold-air
interface (Fig.~\ref{MP_multilayers}(b)). A good quantitative
agreement between the effective medium approximation and exact
transfer-matrix calculations \cite{TorradoNJP13} is obtained.
Observation of significant decrease of the SPP propagation length
$L_{\rm spp}$ for small values of $h$ is attributed to additional
ohmic losses introduced when the ferromagnetic layer is in a
non-negligible overlap with the spatial envelope of the intensity
profile associated with the SPP mode. The influence of a thin
ferromagnetic layer on the skin depth $\delta_{\rm skin}$ is
negligibly small and one can safely use its value for SPPs
propagating at the noble metal-dielectric interface.

The effective medium approximation provides an adequate
description of the dependence of SPP wave vector on the magnetic
field as well. Assuming that the effective magneto-optical tensor
is given by:
\begin{equation}
\label{MOtensor} \widehat{\varepsilon}_{\rm eff}(\pm M)=\left(
\begin{array}{ccc}
\varepsilon_{\rm eff} & 0 & \pm\varepsilon^{xz}_{\rm eff}M \\
0 & \varepsilon_{\rm eff} & 0 \\
\mp\varepsilon^{xz}_{\rm eff}M & 0 & \varepsilon_{\rm eff}
\end{array} \right)\ \,,
\end{equation}
where $M$ denotes the $y$-component of magnetization vector
normalized to its saturation value and the nondiagonal components
are given by
\begin{equation}
\label{e_xz_effective} \varepsilon^{xz}_{\rm
eff}=\frac{1}{\delta_{\rm
skin}}\int_0^{\infty}\varepsilon^{xz}(z){\rm e}^{-z/\delta_{\rm
skin}}dz\,,
\end{equation}
we obtain the following dispersion relation for SPPs
\cite{BelotelovJOSAB2009}:
\begin{equation}
\label{SP dispersion_eff} k_{\rm spp}(\pm M)=
k_0\sqrt{\frac{\varepsilon_{\rm eff}}{1+\varepsilon_{\rm
eff}}}\Big (1\pm\frac{i\varepsilon^{xz}_{\rm eff}M
}{(1-\varepsilon_{\rm eff}^2)\sqrt{\varepsilon_{\rm eff}}}\Big
)\,.
\end{equation}

For a special case of a ferromagnet (FM) layer with thickness
$h_1$ sandwiched between two layers of a noble metal, the
effective medium approximation for the non-diagonal dielectric
susceptibility component gives
\begin{equation}
\label{e_xz_effective} \varepsilon^{xz}_{\rm
eff}\simeq\frac{h_1}{\delta_{\rm skin}}\varepsilon_{\rm
(FM)}^{xz}{\rm e}^{-h/\delta_{\rm skin}}\,
\end{equation}
resulting in the following magneto-plasmonic modulation of SPP
wave vector $\Delta k_{\rm mp}$:
\begin{equation}
\label{ThinFilmFormula} \Delta k_{\rm mp}\simeq i\varepsilon_{\rm
(FM)}^{xz}M\frac{2h_1 \varepsilon_{\rm eff}k_0^2}
{(1+\varepsilon_{\rm eff})(1 -\varepsilon_{\rm eff}^2)}{\rm
e}^{-h/\delta_{skin}} \,,
\end{equation}
where we have introduced the magnetic modulation of SPP wave
vector $2\Delta k_{\rm mp}=k_{\rm spp}(+M)-k_{\rm spp}(-M)$ and
used the expression for SPP skin depth
$\delta_{skin}=\frac{1}{2k_0}{\rm
Im}\frac{\sqrt{1+\varepsilon_{\rm eff}}}{\varepsilon_{\rm eff}}$.
It is remarkable that this result is identical to the analytical
expression obtained from transfer-matrix calculations for the
limit of an infinitely thin ferromagnetic layer
\cite{Temnov10NPhoton4,TorradoNJP13}.

\begin{figure}[h]
  \includegraphics[width=8cm]{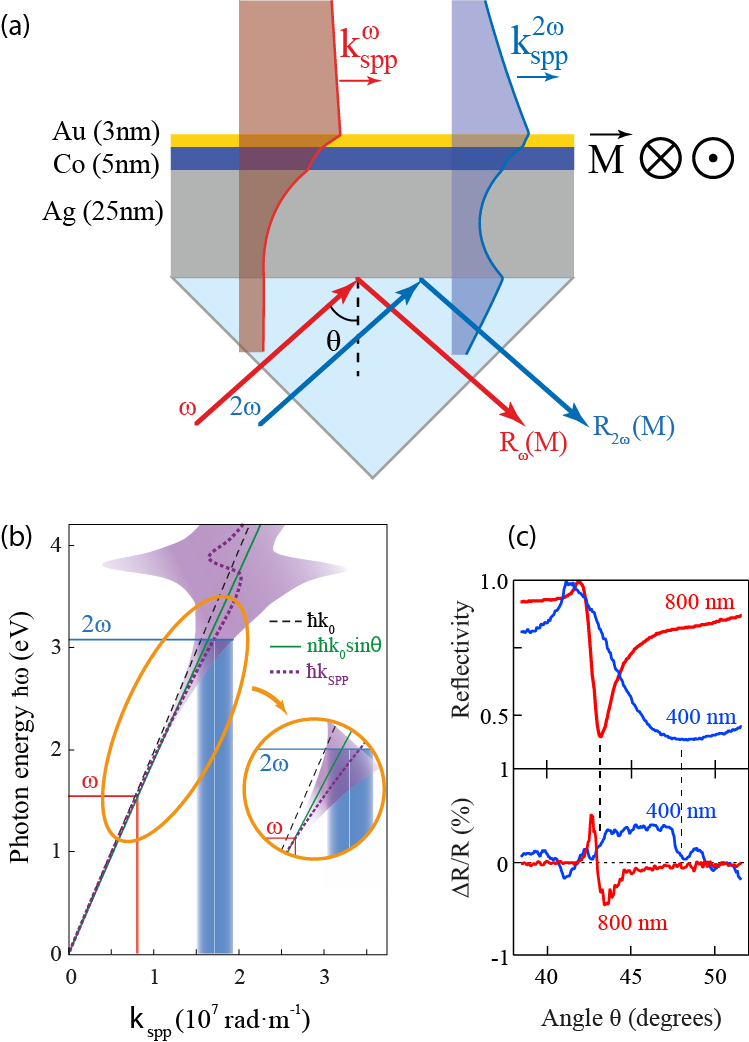}
    \caption{(a) Surface plasmons in thin Au/Co/Ag multilayer
    structures: the spatial distribution of the squared tangential
    projection of SPPs-electric field at the fundamental $\omega$ and
    double $2\omega$ frequencies at their resonant angles.
    (b) Dispersion of the SPP in the Au/Co/Ag trilayer under study.
    Black dash and green solid lines represent the photon dispersion in vacuum and in glass, respectively. Thick dot purple line is the calculated SPP
    dispersion, its linewidth is shown with the purple background area.
    The inset illustrates the possibility of a simultaneous excitation
    of the SPPs at both frequencies $\omega$ and $2\omega$. (c-d) Linear-optical reflectivity $R$ and
    magnetization-induced reflectivity variations $\Delta R/R=[R(+M)-R(-M)]/[R(+M)-R(-M)]$ for the excitation with the 800~nm and 400~nm wavelength. Figure adopted with permission from Ref.~\cite{RazdolskiACS16}.}
    \label{LinearMP_Kretschmann}
\end{figure}

The developed approach appears to be useful not only in the
interferometric measurements with SPPs, but also in the
conventional Kretschmann geometry. In a Au/Co/Ag multilayer the
SPP resonance can be excited at both 800~nm and 400~nm
wavelengths, Fig.~\ref{LinearMP_Kretschmann}(a). The SPP
dispersion in Fig.~\ref{LinearMP_Kretschmann}(b) is displayed by
the shaded area where the width is proportional to SPP damping
(imaginary part of SPP wave vector), which strongly increases when
approaching the blue part of the visible spectral range. Whereas
SPP dispersion results in the shift of the SPP resonance angle
$\theta_0$ for 400~nm light (as compared to 800~nm), the increased
damping makes the Kretschmann reflectivity dip much broader
(Fig.~\ref{LinearMP_Kretschmann}(c)).

The effect of the external magnetic field is the same for both
frequencies and leads to a small shift of the resonance angles
upon magnetization reversal. The positions of the Kretschmann
reflectivity minima precisely correspond to the zeros on
magneto-plasmonic modulation curves, confirming the physical
interpretation suggesting that the magneto-plasmonic modulation of
SPP wave vector should result in an angular shift of the
reflectivity minima.

Here we would like to note that the reflectivity minimum in
Kretschmann configuration cannot always be attributed to the
excitation of SPPs. Depending on the direction of the energy flow
in a thin metal film, i.e. from bottom to the top (Kretschmann
configuration) or from top to the bottom (plasmonic
interferometry, Otto configuration), the dispersion of surface
electromagnetic waves obeying the boundary conditions may be
different. The difference between the conventional SPP modes and
the so-called perfectly absorbed (PA) mode has been experimentally
observed only very recently \cite{FoleySciRep15}. In the
forthcoming discussion of the nonlinear magneto-plasmonic response
we leave the question about the nature of the mode at the
frequency $2\omega$, i.e. whether SPP or PA open.

As we are going to show in the forthcoming sections, the effective
medium approximation for SPPs, which was originally introduced in
the linear magneto-plasmonics, appears to be particularly useful
in description of nonlinear magneto-plasmonics and
acousto-plasmonics.

\subsection{Nonlinear magneto-plasmonics}

Nonlinear magneto-plasmonics is a relatively young field of
magneto-photonics, which utilizes SPP excitations for tailoring
the magnetization-induced contributions to the nonlinear-optical
response of systems. The motivation for the emergence of this
field of research is essentially twofold. First, as compared with
nonlinear plasmonics
\cite{Pacifici08NPhot1,MacDonald08NPhot3,KauranenZayats}, magnetic
field conveniently offers a universal tool for controlling the
nonlinear-optical response. Second, nonlinear magneto-optical
effects are much stronger as compared to their linear
counterparts, where magnetization-induced effects stay relatively
small despite of being enhanced by SPPs
\cite{LupkeReview15,RazdolskiACS16}.

As it has been discussed in the previous section, in linear
magneto-plasmonics the magnetic effects in the transversal Kerr
geometry (magnetization is perpendicular to the incidence plane)
are usually ascribed to the magnetization-induced modulation of
SPP wave vector $k_{\rm spp}$. However, in the nonlinear
magneto-optics the response is more complex. In what follows, we
shall restrict our considerations of nonlinear optical effects to
second harmonic generation (SHG) only, although the general
principles of nonlinear magneto-plasmonics also apply to other
effects such as difference frequency (often in THz domain), third
harmonic generation, etc. Here we focus on the advantages of the
SPP-induced second harmonic generation in magnetic media,
including both enhancement of the magnetic effects and unraveling
new SPP-assisted mechanisms to control optical nonlinearities.

One of the basic nonlinear-optical processes, SHG is governed by
the second-order nonlinear-optical susceptibility tensor
$\chi^{(2)}$. Various mechanisms of the anharmonicity in the
optical response to electromagnetic field $E(\omega)$ lead to the
emergence of nonlinear polarization, which, at the lowest second
order, occurs at the double frequency $2\omega$
\cite{RaschkeChapter7}:

\begin{equation}
P_i(2\omega)=\chi^{(2)}_{ijk}(-2\omega;\omega,\omega):E_j(\omega)E_k(\omega)\,.
\end{equation}
This polarization then emits an electromagnetic wave. One of the
key properties of the $\chi^{(2)}$ tensor is its strong
sensitivity to the inversion symmetry. It can be shown that
dipolar $\chi^{(2)}$ vanishes in the centrosymmetric media such
as, for instance, bulk metals. In such systems efficient SHG is
generated predominantly at the interfaces, where the inversion
symmetry is broken. The renowned SHG sensitivity to the
interfacial properties of media is advantageous for plasmonics
with SPPs. Under resonant excitation conditions in Kretschmann
configuration SPPs boost up the local electromagnetic fields and
can strongly enhance the SHG output \cite{SimonPRL74}. In certain
cases it is also convenient to take into account the influence of
SPPs in the form of a resonant contribution to the
$\chi^{(2)}$-tensor \cite{RaschkeChapter7,Heinz}. Importantly,
since $\chi^{(2)}(-2\omega;\omega,\omega)$ is sensitive to
resonances (of various nature) not only at the fundamental
$\omega$, but also at the double frequency $2\omega$, SPPs at both
of these frequencies contribute to the SHG output
\cite{MartiniPRL76,PalombaPRL08}.

\begin{figure}[h]
  \includegraphics[width=0.95\columnwidth]{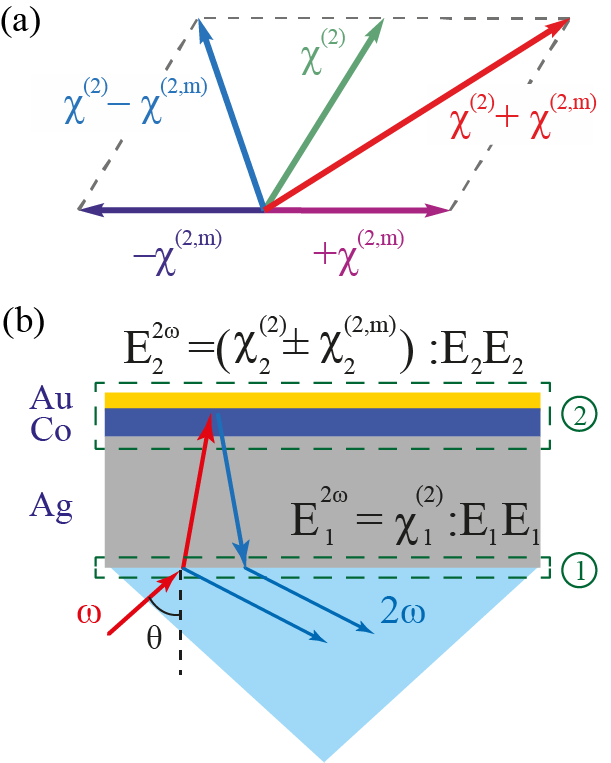}
    \caption{
    (a) Illustration of the origin of the inequality of the SHG intensities for the opposite direction of the magnetization $M$ leading to the non-zero magnetic SHG contrast $\rho_{2\omega}$. The sign of the magnetic contribution to the nonlinear susceptibility $\pm\chi^{(2,m)}$ is flipped when the magnetization is reversed.
    (b) Magnetic SHG generation in a multilayer structure excited in Kretschmann geometry: SHG sources at the two interfaces, bottom (1, glass/Ag) and top (2, air/Au/Co/Ag) within the effective interface approach.
    }
    \label{mSHG}
\end{figure}

Within this formalism, the magnetization-induced modulation of the
SHG output is described in the following way. In magnetized media,
the magnetization-dependent $\chi^{(2)}$ tensor can be represented
as a sum of the non-magnetic (crystallographic) $\chi^{(2)}$ and
magnetization-induced $\pm\chi^{(2,m)}$ contributions
\cite{rupinpan}. The latter changes sign upon reversing the
magnetization, whereas the interference of these two contributions
results in a magnetization-induced modulation of the SHG intensity
(Fig.~\ref{mSHG}(a)). The magnitude of the effect can be
conveniently characterized by the so-called magnetic SHG (mSHG)
contrast $\rho_{2\omega}$:

\begin{equation}
\rho_{2\omega}=\frac{I_{2\omega}(+M)-I_{2\omega}(-M)}{I_{2\omega}(+M)+I_{2\omega}(-M)},
\end{equation}
where $I_{2\omega}(+M)$ and $I_{2\omega}(-M)$ are the SHG
intensities measured for the two opposite directions of
magnetization. It should be noted that in a complex multilayer
system, where multiple SHG sources are located at each interface,
the rigorous description of the SHG output becomes increasingly
complicated. Instead, a rather simple approach of the effective
interface can be introduced, where a stack of thin metallic layers
is replaced by a single SHG source (Fig.~\ref{mSHG}(b)). Within
this approach, the role of SPP excitation in altering the
magnetization-induced effects in SHG can be considered twofold.
Upon such an excitation, either the relative magnitude of the
$\chi^{(2)}$ contributions or the relative phase between them can
be modulated. The former, realized by Krutyanskiy {\it et al.}
\cite{KrutyanskiyPRB13} relies on the tensorial nature of
$\chi^{(2)}$. Here, the crystallographic and magnetic SHG
contributions are provided by the different projections of the
fundamental field $E(\omega)$. These projections experience
unequal SPP-assisted enhancement resulting in a modulation of the
magnetization-induced SHG output. The second option
\cite{RazdolskiPRB13,KrutyanskiyPRB15} highlighted the importance
of the SPP-induced variations of the phase of the electromagnetic
field. The physical origin of such phase modulations beyond the
usual field enhancement remains largely unexplored.

A number of systems has been successfully investigated with
nonlinear magneto-plasmonics. We note that nonlinear
magneto-optics of nanoparticles and planar nanostructures
\cite{MurzinaSurfSci01,AktsipetrovJOSAB05,KolmychekSSPh09,ValevACS11}
supporting localized surface plasmon resonances are beyond the
scope of this work. The most straightforward approach utilizing
SPP excitation at the interface of a ferromagnetic (Fe, Co, Ni)
rather than conventional plasmonic (Al, Ag, Au, Co) metals faces a
few challenges. Apart from dealing with overdamped plasmonic
excitations ($L_{\rm spp}<\lambda_{\rm spp}$), these metals
possess a relatively low second-order nonlinearity usually
attributed to the localized $d$-band electrons. Nevertheless,
magnetic SHG studies both on periodically perforated films
\cite{RazdolskiPRB13} and in Kretschmann geometry
\cite{ZhengSciRep14} demonstrated significant SPP-assisted
modulation of the magnetic contrast $\rho_{2\omega}$. In the
systems discussed above, the incident electromagnetic wave excited
SPPs at the fundamental frequency $\omega$ and the plasmonic
enhancement of the local electromagnetic field $E(\omega)$ boosted
the efficiency of the nonlinear-optical conversion.

Multilayer structures consisting of a ferromagnetic layer
sandwiched between the two layers of noble metals offer a
considerable expansion of opportunities in nonlinear
magneto-plasmonics. For example, photons initially up-converted
(frequency-doubled) at the interfaces can excite SPPs at $2\omega$
and thus contribute to the enhancement of the SHG output
\cite{PalombaPRL08,GrossePRL12}. This mechanism requires the
sample to support propagating SPP modes at the double frequency
$2\omega$ and imposes additional requirements on the system. It
can be especially challenging in the case of SPP excitation on
periodically corrugated surfaces, where the spatial periodicity
was designed to achieve resonant conditions for SPP excitation at
the fundamental frequency $\omega$ only. Another impairing factor
is related to increased optical losses caused by the spectral
overlap of SPP at $2\omega$ with interband transitions, which is
the case for Au at photon energies above 2.4~eV. This shifts the
focus of attention toward other plasmonic metals such as Al or Ag
which are capable of sustaining SPPs across the whole visible
spectral range.

In the following we discuss a showcase of the proposed nonlinear
magneto-plasmonic mechanism where a significant improvement of the
magnetization-induced modulation reaching up to 33\% has been
achieved. We consider a thin gold/cobalt/silver trilayer grown on
a glass substrate by means of the magnetron sputtering. A
5~nm-thin magneto-optically active layer of ferromagnetic cobalt
was protected from oxidation by a 3~nm-thin layer of gold. A
25~nm-thick silver layer acted as the main constituent in this
hybrid plasmonic nanostructure, which was excited by collimated
100~fs short laser pulses through the glass prism. The reflected
SHG intensity was recorded as a function of the incidence angle
$\theta$ for the two opposite directions of magnetization in
cobalt in the transverse Kerr geometry (see Fig.~8(a)).

Due to the inevitable dispersion, the SPP excitations at
fundamental and double frequencies in Kretschmann geometry occur
at slightly different angles. However, nonlinear-optical
considerations discussed above suggest a possibility of nonlinear
phase-matching between the second harmonic SPP at the gold-air
interface with the $k$-vector $k^{2\omega}_{\rm spp}$ and the
excitation source at the silver-glass interface characterized by
the in-plane component of the $k$-vector
$n(\omega)k_0(\omega)\sin\theta$. This phase-matching occurs at an
angle $\theta_{\rm nl}$ given by:
\begin{equation}\label{PhaseMatching}
k_{\rm spp}^{2\omega}=2n(\omega)k_0(\omega)\sin\theta_{\rm nl}\,,
\end{equation}
Being just one of several possible SPP frequency conversion
pathways \cite{GrossePRL12,HeckmannOpEx13}, this phase-matching
condition is of paramount importance as it determines the resonant
SPP-induced enhancement of the nonlinear susceptibility
$\chi^{(2)}=\chi^{(2)}_{\rm nr}+\chi^{(2)}_{\rm res}(\theta)$ with
SPP-mediated resonant contribution \cite{RaschkeChapter7,Heinz}:
\begin{equation}\label{chires}
\chi^{(2)}_{\rm res}(\theta)\propto\frac{1}{\theta-\theta_{\rm
nl}+i\Gamma}\,
\end{equation}
with $\Gamma={\rm Im}[k^{2\omega}_{\rm
spp}]/(k_0(\omega)n(\omega))$.

\begin{figure}[t]
    \begin{centering}
    \includegraphics[width=6cm]{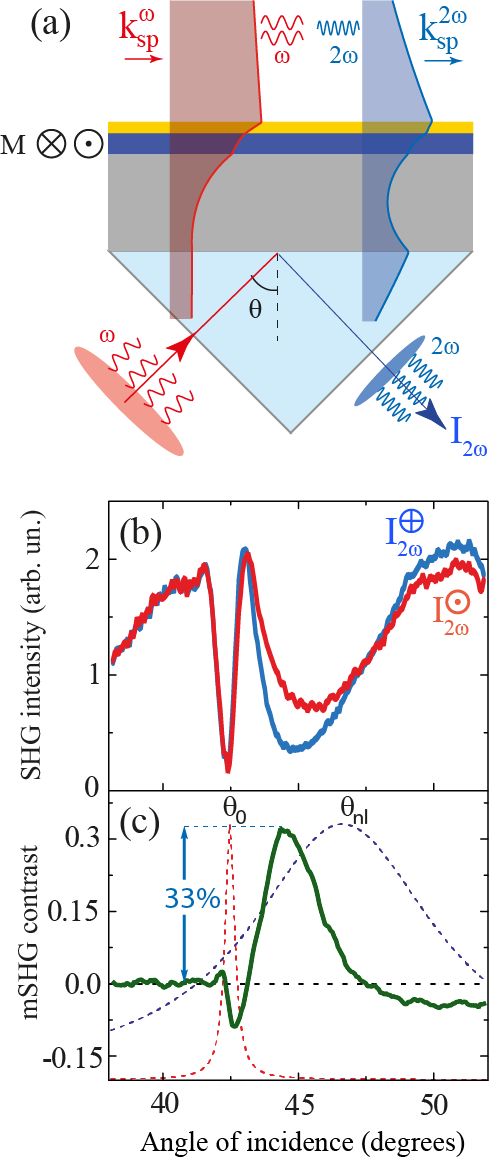}
    \caption{
    (a) Schematic of the SPP-induced magnetic SHG in Kretschmann geometry. The shaded areas represent the calculated distributions of the square of the SPP electric field $|E_z|^2$. A noticeable field enhancement at the air/Au interface indicates that the SPPs at both $\omega$ and $2\omega$ are excited. Angular dependence of (b) the SHG intensity for the two opposite directions of magnetization in Co (red and blue) and
    (c) mSHG magnetic contrast (green). Red and blue dashed lines represent the SPP resonance lines for the 800~nm pump wavelength (described by Eq.~(\ref{LinearSPRcondition})) and the nonlinear (SHG) excitation at 400~nm wavelength (according to Eq.~(\ref{PhaseMatching}), respectively. Note that the maximum of mSHG contrast is located between fundamental and the nonlinear (SHG) SPP resonances.
    }
    \label{mSHGdata800}
    \end{centering}
\end{figure}

Whereas the linear reflectivity at both frequencies $\omega$ and
$2\omega$ shows only small modulations of the order of 1\%, as
discussed in the previous section (Fig.~6(c)), the angular
dependence of the nonlinear SHG spectra in
Fig.~\ref{mSHGdata800}(b) displays drastic changes. Contrary to
the previously reported results on a gold film
\cite{PalombaPRL08}, the angular positions of SPP resonances for
the fundamental and SHG frequencies in our multilayer structure
correspond to the pronounced minima in the SHG intensity. A strong
dependence of the total SHG intensity on the magnetization
direction $I_{2\omega}(\pm M)$ is quantified by the magnetic SHG
contrast $\rho_{2\omega}$ shown in Fig.~\ref{mSHGdata800}(c).

It is seen that the largest magnetisation-induced modulation of
the SHG intensity is accompanied by the SPP excitation at the SHG
frequency and not the fundamental one. Angular dependence of mSHG
contrast displays a large  modulation amplitude reaching 33~$\%$
at $\theta=44$~degrees. The data for other excitation wavelengths
can be found elsewhere \cite{RazdolskiACS16}. For the shortest
wavelength (760 nm) the mSHG maximum is only about 20\%, since in
this case the SPP damping at the SHG frequency becomes so large
that the system approaches the region with the non-propagating
(overdamped) SPPs.

Note that the mSHG contrast at the fundamental SPP resonance is
barely reaching 10\%. This observation is in line with the most
recent results by Zheng {\it et al.} \cite{ZhengSciRep14}, who
reported similar values of the mSHG contrast on a 10~nm-thin iron
film on glass, as well as with the results by Pavlov {\it et al.}
\cite{PavlovAPL99} obtained on Au/Co/Au multilayer, the structures
not supporting SPPs at the SHG frequency. This fact, along with
the dispersive shift of mSHG maximum reinforces our conclusion
that the 33~$\%$ large mSHG contrast (which is equivalent to the
increase of the SHG intensity by a factor of 2 upon magnetization
reversal) is dominated by the nonlinear SPP resonance at the SHG
frequency.

Following Palomba and Novotny \cite{PalombaPRL08}, the complex
angular dependence of the SHG intensity from a thin gold film on
glass is explained by an interference of two contributions coming
from the metal-air and metal-glass interfaces. In our case
(Fig.~\ref{mSHG}(b)) the silver-glass interface acts as a source
of the nonmagnetic SHG $\vec{E}^{2\omega}_{1}$, and the upper part
consisting of Au and Co layers is assumed to generate the electric
field $\vec{E}^{2\omega}_{2}$ containing both magnetic
$\chi^{(2m)}$ and non-magnetic $\chi^{(2)}$ contributions. Thus
the total SHG intensity $I_{2\omega}$ is described by:
\begin{eqnarray}\label{intensity}
 I_{2\omega}&\propto& |\vec{E}^{2\omega}_1+\vec{E}^{2\omega}_2\pm
 \vec{E}^{2\omega}_{2m}|=\\
&=&|\chi^{(2)}_1:\vec{E}_1 \vec{E}_1 + (\chi^{(2)}_2\pm
\chi^{(2m)}_2):\vec{E}_{2} \vec{E}_{2}|^2\,\nonumber
\end{eqnarray}
Here the complex tensor components $\chi^{(2)}_1$ and
$\chi^{(2)}_2\pm \chi^{(2m)}_2$ represent the effective optical
nonlinearities at both interfaces. Owing to the resonant
$\chi^{(2)}_{\rm res}$ contribution, the SHG field enhanced at the
top interface destructively interferes with the one generated at
the bottom interfaces, which explains the experimentally observed
SHG intensity minima. Within this approach, using
Eq.~(\ref{intensity}) and assuming the resonant $\chi^{(2)}_{\rm
res}$ given by Eq.~(\ref{chires}) we were able to fit the
experimental angular spectra of both SHG intensity and magnetic
contrast \cite{RazdolskiACS16}. The frequency dependence over the
dispersive SPP spectral range revealed that the maximum value of
33\% mSHG contrast remained constant whereas its angular width
decreased as the nonlinear SPP resonance became narrower ($\Gamma$
decreased) and shifted towards the fundamental one. As such, the
dominant role of the nonlinear SPP resonance in Kretschmann
geometry, which is responsible for the increase of the mSHG
magnetic contrast, was identified.

In conclusion, the nonlinear magneto-plasmonics offers a strong
enhancement of the magnetization-induced effects as compared to
its linear counterpart. In the given spectral range one would
expect to reach further enhancement of mSHG contrast by
systematically varying the individual thicknesses in this trilayer
structure.

\section{Acousto-plasmonics}
\subsection{Linear acousto-plasmonics}

In previous sections we have investigated stationary properties of
metal-ferromagnet multilayer structures. Femtosecond light pulses
were used to achieve high peak intensities of electromagnetic
radiation facilitating efficient nonlinear optical frequency
conversion. However, irradiation of metallic samples with
femtosecond optical pulses is known to trigger the complex
dynamics of electronic, acoustic and magnetic excitations
\cite{Beaurepaire96PRL76,vanKampen2002,Eschenlohr2013,ShalagatskyiArxiv2015}.
In this section we will discuss transient effects of ultrashort
acoustic pulses on SPPs.

Among different mechanisms of acoustic generation the so-called
thermo-elastic mechanism is the most common one
\cite{Thomsen86PRB34,MatsudaUltrasonics2015}. Absorption of light
leads to fast rise of temperature on a sub-picosecond time scale
and the build-up of thermo-elastic stress followed by thermal
expansion. This process results in the emission of coherent
acoustic pulses with the shape resembling the initial energy
deposition profile with a characteristic spatial scale
$\delta_{\rm heat}$. The acoustic pulse duration is given by
$\delta_{\rm heat}/c_s$, where $c_{\rm s}=\sqrt{C_2/\rho}$ denotes
the longitudinal speed of sound, $\rho$ is material density and
$C_2$ is determined by the elastic tensor and depends on the
acoustic propagation direction \cite{Hao01PRB64}.

The heat deposition depth $\delta_{\rm heat}$ is determined not
only by the optical skin depth, but also by the diffusion depth of
laser-excited hot electrons during the thermalization time with
initially cold lattice. This depth is given by $\delta_{\rm
hot}\sim\sqrt{\kappa/g}$, where $\kappa$ and $g$ denote the
thermal conductivity and electron-phonon coupling constant in the
material, respectively \cite{Tas94PRB49}. In plasmonic metals
characterized by relatively high thermal conductivity and weak
electron-phonon coupling \cite{DelFatti00PRB61} the heat
deposition depth $\delta_{\rm heat}\simeq\delta_{\rm
hot}\sim$100~nm largely exceeds the optical skin depth resulting
in the generation of relatively long acoustic pulses with the
duration of a few tens of picoseconds \cite{GusevPRB1998}.

For noble metal films of thickness smaller than $\delta_{\rm
heat}$, the temperature distribution is spatially homogeneous and
the thermo-elastic dynamics can be reduced to breathing motion of
the entire film consisting of the alternating expansion and
contraction. This behaviour was evidenced in femtosecond
time-resolved pump-probe experiments in Kretschmann configuration
\cite{vanExter88PRL60,Wang2007OL32,Wang2007PRB75}.

Hot electron diffusion in ferromagnetic metals appears to be less
efficient ($\delta_{\rm hot}\sim\delta_{skin}\sim$10~nm) and
results in the overall heat penetration depth $\delta_{\rm
heat}\sim$15-20~nm \cite{Saito03PRB67,TemnovNatureComm2013}. This
short heat penetration depth in combination with a larger speed of
sound suggests the generation of large-amplitude ultrashort
acoustic pulses in ferromagnetic transition metals like Ni, Co and
Fe. However, SPPs do not propagate on ferromagnet surfaces thereby
rendering the plasmonic detection schemes extremely challenging.
Up to now, there are no acousto-plasmonic investigations on thin
ferromagnetic films.

\begin{figure}[h]
  \includegraphics[width=8 cm]{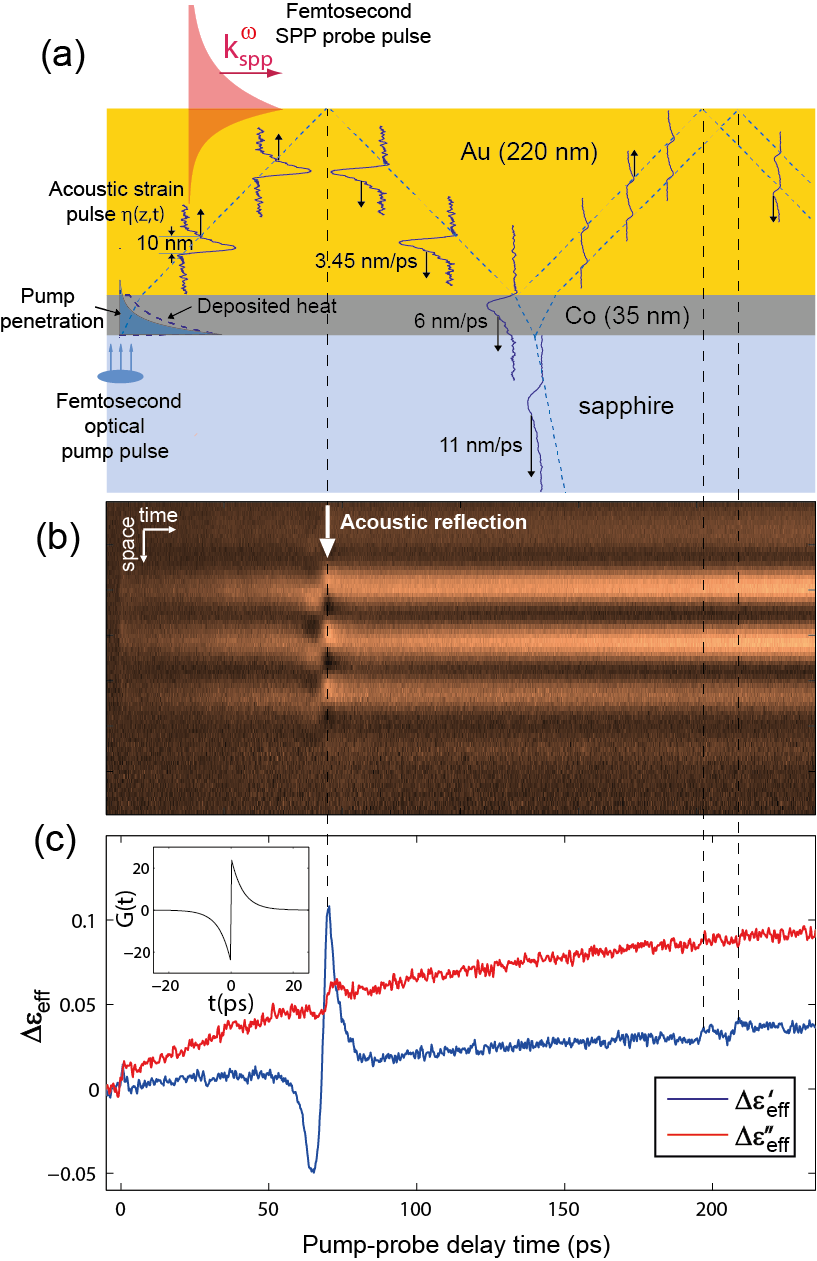}
    \caption{(a) In hybrid acousto-plasmonic structures the thermal expansion
of cobalt transducer excited by a femtosecond pump pulse (fluence
7~mJ/cm$^2$) at time zero launches an ultrashort acoustic pulse
$\eta(z,t)$ propagating through (111) textured  gold layer at the
speed of sound $c_s=3.45$~km/s. After approximately 70~ps, the
acoustic reflection from gold-air interface changes the wave
vector of a time-delayed ultrashort SPP probe pulse generating the
acousto-plasmonic pump-probe interferogram (b) in a tilted
slit-groove arrangement. The 10~nm wide acoustic strain pulse in
(a) has a duration of 3~ps and is reconstructed from the dynamics
of the time-dependent effective dielectric function
$\Delta\varepsilon_{\rm eff}(t)$ in (c). The inset shows the
acousto-plasmonic response function $G(t)$. Measurements are
performed by time-resolved SPP interferometry
\cite{SPInterferometry09OE,TemnovNatureComm2013}}
 \label{AcoustoPlasm1}
\end{figure}

The use of hybrid (noble metal)-ferromagnet bilayer structures on
a dielectric substrate allows for the generation of ultrashort
acoustic pulses in the ferromagnetic transducer and their
plasmonic detection  at the (noble metal)/air interface, see
Fig.~\ref{AcoustoPlasm1}(a).

As the duration of the acoustic pulses is conserved upon the
transmission from one elastic medium into another one, acoustic
pulses generated in cobalt are spatially compressed by a factor of
1.8 (ratio of sound velocities in cobalt and gold) down to 10~nm
upon injection in softer gold \cite{TemnovNatureComm2013}.

Femtosecond time-resolved plasmonic interferometry
\cite{SPInterferometry09OE} measures transient phase shifts of the
plasmonic interferogram at the gold-air interface
(Fig.~\ref{AcoustoPlasm1}(b)) caused by thermal and acoustic
effects triggered by the absorption of an intense femtosecond pump
pulse in cobalt layer. The reconstructed pump-induced modulation
$\Delta\varepsilon_{\rm eff}(t)=\varepsilon_{\rm
eff}(t)-\varepsilon_{\rm Au}$ in Fig.~\ref{AcoustoPlasm1}(c)
displays the strong acoustic modulation capturing the reflection
of the acoustic pulse $\eta(z,t)$ from the gold-air interface at
70~ps acoustic delay superimposed on a slowly varying thermal
background (which will be neglected in the following discussion).
The much weaker acousto-plasmonic signals observed at 197~ps and
208~ps are caused by the secondary 10\% acoustic reflections from
the gold-cobalt and cobalt-sapphire interfaces, respectively, and
indicate  a good acoustic impedance matching at these interfaces.

The quantitative reconstruction of the acoustic pulse $\eta (z,t)$
propagating in gold relies on the effective medium approximation.
The compressional acoustic pulse in gold $\eta(z,t)=(n_{\rm
i}(z,t)-n_{\rm i}^0)/n_{\rm i}^0$ creates a layer of higher ion
density $n_{\rm i}(z,t)>n_{\rm i}^0$, which moves at the sound
velocity $c_{\rm s}$=3.45~nm/ps in gold in the (111) direction.
Since the stationary charge separation between electrons and ions
in a metal occurs only within the Thomas-Fermi radius $r_{\rm
TF}\sim 10^{-3}$~nm, the spatial profile of electron (charge)
density $n_{\rm e}(z,t)$ follows the ionic one: $n_{\rm
e}(z,t)=n_{\rm i}(z,t)$. Evaluated at the probe photon energy of
1.55~eV, the dielectric function of gold, $\varepsilon_{\rm
Au}=\varepsilon^{'}+i\varepsilon^{''}=-24.8+1.5i$, is dominated by
the free-carrier contribution with $\varepsilon^{'}\simeq
-\omega^2_{\rm p}/\omega^2\propto -n_{\rm e}$. An ultrashort
acoustic strain pulse creates a time-dependent spatial profile of
the dielectric function $\varepsilon^{'}(z,t)=
\varepsilon^{'}[1+\eta(z,t)]$ inside the metal, which modulates
the SPP wave vector $k_{\rm spp}(t)=k_0\sqrt{\varepsilon_{\rm
eff}(t)/(1+\varepsilon_{\rm eff})}$, when the strain pulse arrives
within the SPP skin depth $\delta_{\rm skin}=13$~nm at the
gold-air interface:
\begin{equation}
\label{AcPlasmEquation} \varepsilon_{\rm
eff}(t)=\frac{\varepsilon_{\rm Au}}{\delta_{\rm skin}}
\int_0^{\infty}[1+\eta(z,t)]{\rm exp}(-z/\delta_{\rm skin})dz \,.
\end{equation}
Using the explicit equation for the acoustic reflection at the
(free) gold-air interface, $\eta(z,t)=\eta_0(t+z/c_{\rm
s})-\eta_0(t-z/c_{\rm s})$, the integral over space can be
converted into the integral over time. The real part of the
acoustic modulation
\begin{equation}
\label{Epsilon1} \Delta\varepsilon^{'}_{\rm
eff}(\tau)=\frac{|\varepsilon_{\rm Au}|}{\tau_{\rm skin}}
\int_{-\infty}^{\infty}\eta_0(t)G(\tau-t)dt \,
\end{equation}
of the effective dielectric function can be used to reconstruct
the acoustic pulse $\eta_0(t)$. Here $\tau_{\rm skin}=\delta_{\rm
skin}/c_s$=3.8~ps denotes the acoustic travel time through the
skin depth in gold (at 800~nm wavelength) and the
acousto-plasmonic response function $G(t)$, defined as
\begin{equation}
\label{ResponseFunction} G(t)={\rm exp}(-|t|/\tau_{\rm skin}){\rm
sign}(t)\,,
\end{equation}
is shown in the inset in Fig.~\ref{AcoustoPlasm1}(c). Application
of the  Fourier-based algorithm \cite{MankeAPL2013} to
$\Delta\varepsilon^{'}_{\rm eff}(\tau)$ allows to accurately
reconstruct the acoustic pulse shape $\eta_0(t)$ with duration of
3~ps and spatial extent of 10~nm, as shown in
Fig.~\ref{AcoustoPlasm1}(a). It is remarkable that plasmonic
measurements provide the detailed shape of an acoustic pulse with
the spatial dimensions smaller than the skin depth $\delta_{\rm
skin}$.

Under certain conditions such shorter-than-skin-depth acoustic
pulses can be also measured by simple time-resolved reflectivity
measurements at the  metal-dielectric interface, as shown for the
gold-air interface at 400~nm optical probe wavelength
\cite{MankeAPL2013}.

\subsection{Nonlinear acousto-plasmonics}

\begin{figure}[h]
  \includegraphics[width=8 cm]{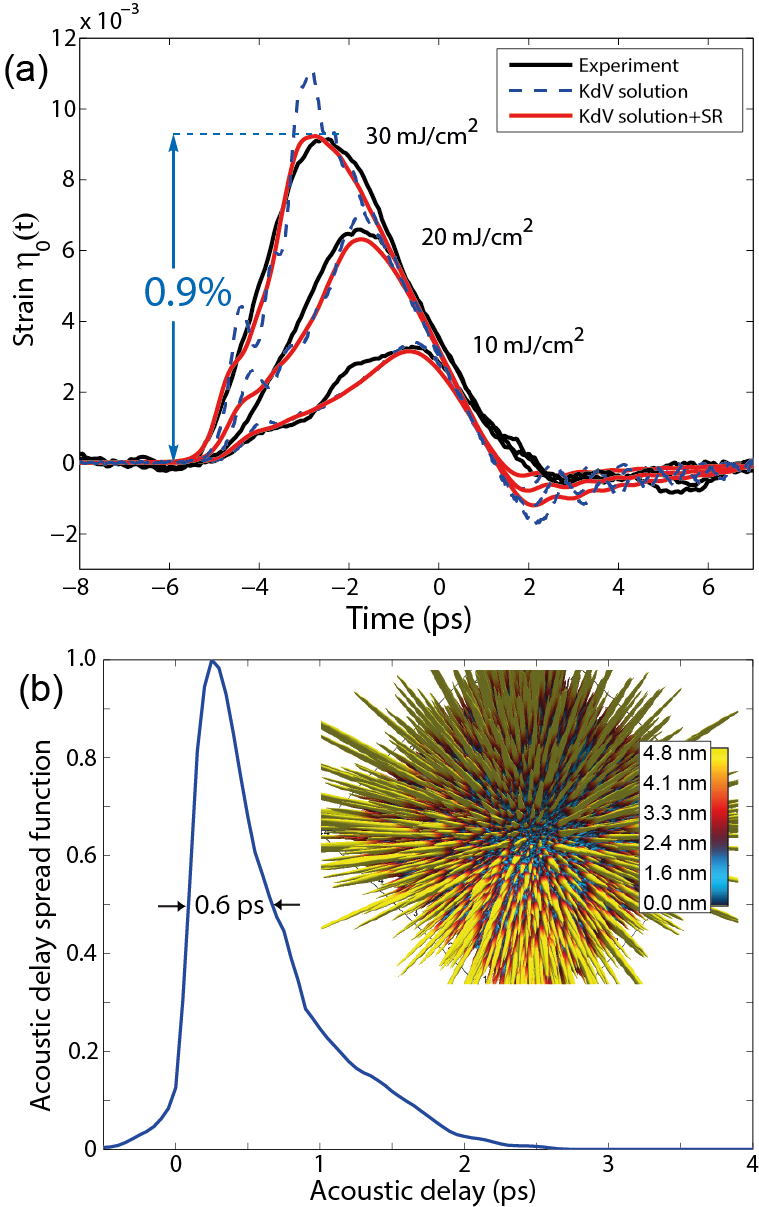}
    \caption{(a) The amplitude of acoustic pulses grows with the pump fluence, reaching nearly 1$\%$ whereas their shape changes. These shape changes are reversible and can be quantitatively described by the nonlinear acoustic propagation effects through a 220~nm thin gold layer.
    (b) Examination of the surface roughness (SR) by an atomic force microscopy (AFM) allows for the measurement of the topography of the gold-air interface. The nano-scale surface roughness (see the 3D-map of 5$\times5~\mu$m$^2$ AFM scan in the inset) introduces substantial spread of the acoustic arrival times at the gold-air interface, quantified by the acoustic delay spread function.
    Figure (a) adopted with permission from ref.~\cite{TemnovNatureComm2013}.}
 \label{AcoustoPlasm2}
\end{figure}

The most striking advantage of the plasmonic interferometry as
compared to the conventional pump-probe reflectivity analysis, is
that it provides direct measurement of the absolute strain values.
Upon increase of the peak pump fluence up to 30~mJ/cm$^2$, the
acoustic pulses reach very large amplitude of 1$\%$ and also
change their shape, see Fig~\ref{AcoustoPlasm2}. The acoustic
reshaping at high pump fluences becomes more pronounced in thicker
gold films, which allowed us to establish its close relation to
the acoustic propagation effects in gold
\cite{TemnovNatureComm2013}. The peak of the acoustic pulse, where
the density of gold is higher, propagates through the gold layer
at a higher speed of sound leading to the steepening of the
acoustic pulse.

The results of the acousto-plasmonic measurements are found to be
in an excellent quantitative agreement with the solutions of the
non-linear Korteveg-de Vries (KdV) equation
\cite{vanCapelUltrasonics2015}
\begin{equation}
\label{KdV} \frac{\partial\eta}{\partial t}+c_{\rm
s}\frac{\partial\eta}{\partial z}
+\gamma\frac{\partial^3\eta}{\partial z^3}+\frac{C_3}{2\rho
c_s}\eta\frac{\partial\eta}{\partial z}=0 \,.
\end{equation}
These dynamics are governed by the interplay between the acoustic
broadening due to the phonon dispersion $\omega(q)=c_{\rm
s}q-\gamma q^3$ with $\gamma=7.41\times 10^{-18}$~m$^3$/s and
self-steepening and reshaping due to the elastic nonlinearity
$C_3=-2.63\times 10^{12}$~kg/ms$^2$. See Ref.~\cite{Hao01PRB64}
for a definition of $C_3$ and
Ref.~\cite{Behari70JPCSSP3,Hiki66PR144} for linear and higher
order elastic constants in gold; $\rho=19.2$~g/cm$^3$ is the
density of gold. The agreement between the nonlinear theory and
the experiment is achieved by setting a single fit parameter, the
initial heat penetration depth in cobalt to $\delta_{\rm
heat}=20$~nm (corresponding to the initial pulse duration
$\tau_0=3.2$~ps). This value is 50\% larger than the optical skin
depth in cobalt of 13~nm, indicating the importance of the
electronic transport phenomena for establishing the initial
distribution of thermal stress driving the thermo-elastic
generation.

The quantitative agreement between theory and experiment for
samples with different gold thickness demonstrated that nonlinear
reshaping was dominated by acoustic nonlinearities in gold,
whereas a possible dependence of the initial strain shape in
cobalt on pump power played a minor role
\cite{TemnovNatureComm2013}. The amplitude of the acoustic strain
reached peak values $\eta_{\rm max}\simeq1\%$, corresponding to a
compressional stress $dp=c^2_{\rm s}\rho\eta_{\rm max}=2.3$~GPa
before acoustic reflection and negative -2.3~GPa tensile stress
afterwards. Control measurements at lower pump fluences confirmed
that the nonlinear reshaping presented in
Fig.~\ref{AcoustoPlasm2}(a) was fully reversible. At even higher
pump fluences in the range of 50~mJ/cm$^2$, strain pulses with
amplitudes reaching 1.5\% were obtained. However, under such
strong optical pumping consistent with elevating the lattice
temperature in cobalt close to its melting point of 1768~K,
irreversible degradation of the samples was observed.

Intrinsic damping of longitudinal phonons in gold caused by
anharmonic phonon-phonon interactions \cite{Tang11PRB84} becomes
increasingly important for frequencies exceeding 1~THz. However,
in this frequency range and for our experimental geometry, the
effect of nano-scale surface roughness entirely masks possible
contributions due to phonon attenuation.

In order to illustrate this effect we have studied the topography
of the gold-air interface at the position on the sample where
acousto-plasmonic measurements were performed, by means of the
atomic force microscopy (AFM). The inset in
Fig.~\ref{AcoustoPlasm2}(b) shows a 3D representation of surface
as seen by the acoustic pulses: given the tiny 3.45~nm wavelength
of longitudinal acoustic phonons in (111) gold at 1~THz frequency,
the nanoscale surface roughness results in the substantial
distribution of acoustic arrival times at the gold-air interface.
This effect is quantified in the histogram of the acoustic arrival
times at the gold-air interface, which we denote as an acoustic
delay spread function with a characteristic width of 0.6~ps.
Therefore, all possible high-frequency components exceeding 1~THz
forming ultrafast acoustic transients on a sub-picosecond time
scale (such as, for example, acoustic solitons
\cite{Hao01PRB64,vanCapel10PRB81,MarisPRB2011,vanCapelUltrasonics2015})
are smeared out by the convolution with the response function
(compare KdV and KdV+SR curves in Fig.~\ref{AcoustoPlasm2}(a)).

The nonlinearity of acoustic propagation can be used to calibrate
the conventional pump-probe reflectivity measurements, when the
latter are compared with a quantitative experimental technique
capable to measure  the absolute values of strain amplitudes. Both
ultrafast X-ray diffraction \cite{BojahrPRB2012} and
acousto-plasmonic interferometry \cite{TemnovNatureComm2013} do a
decent job. For example, the comparison of the KdV solutions with
the fingerprints of the nonlinear acoustic reshaping observed in
femtosecond pump-probe reflectivity measurements on the same
structures \cite{PezerilOE2014} provides the values for
photo-elastic coefficients $dn/d\eta=2\pm0.7$ and
$dk/d\eta=1\pm0.3$ in gold at 400~nm probe wavelength, where
$n+ik=1.47+1.95i$ denotes the complex index of refraction.
Accurately calibrated time-resolved reflectivity measurements
carry the same physical information while offering considerable
advantage over more complex quantitative techniques based on SPPs
or X-rays.

When thinking in terms of the Kretschmann configuration, the
nonlinearity of the acoustic propagation is unlikely to play a
role due to the much smaller thickness, which typically does not
exceed 50~nm for the entire multilayer structure.
In contrast, the entire excitement about the nonlinear
magneto-plasmonics is triggered by a much larger effect in the
optical nonlinearities as compared to conventional plasmonics. It
would be very instructive to learn if SHG intensity can be
modulated by ultrashort acoustic pulses as well.

Here we illustrate a novel and poorly understood phenomenon of SHG
modulation by acoustic pulses
\cite{ZhaoPRB2011,BykovPRB2015,HuberPRB2015} with an example of
femtosecond pump-probe measurements on a Au/Fe/Au/Fe multilayer
structure grown epitaxially on a MgO(001) substrate. This
multilayer structure has been designed to study ultrafast
transport of spin-polarized hot carriers in gold injected from
laser-excited ferromagnetic iron discussed in
Ref.~\cite{MelnikovPRL2011}.

\begin{figure}
\begin{centering}
  \includegraphics[width=8cm]{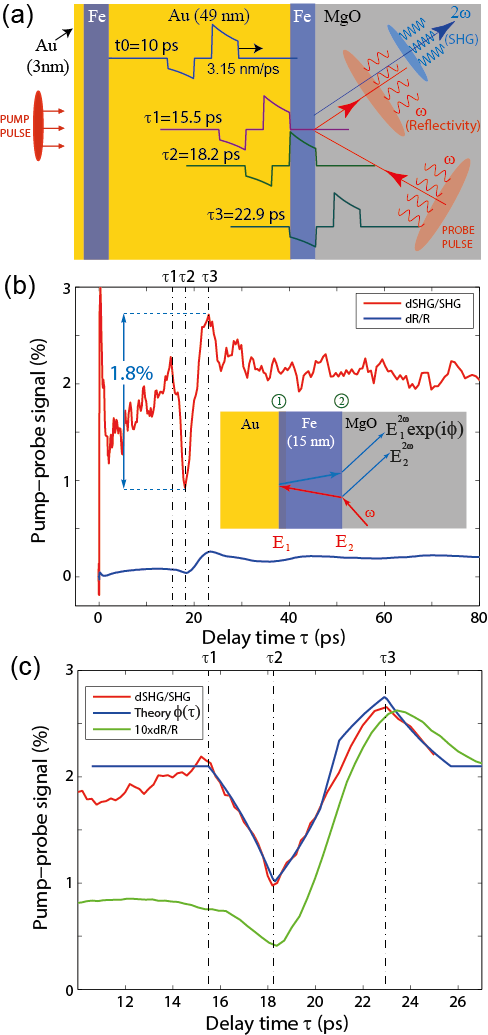}
    \caption{ (a) Femtosecond optical pumping of an epitaxial (3~nm)Au/(16~nm)Fe/(49~nm)Au/(15~nm)Fe/MgO multilayer structure leads to the generation of a bipolar acoustic pulse propagating across the sample.
    (b)  When passing through the second Fe layer, the acoustic pulse changes both linear reflectivity and SHG response to the ultrashort probe pulses. The inset shows a possible mechanism of SHG generation via two interfering contributions $E^{2\omega}_1$ and $E^{2\omega}_2$ from Au/Fe and Fe/MgO interfaces.
    (c) A high-resolution temporal scan shows a remarkable similarity in the dynamics of a strain-induced modulation of the optical phase $\phi(\tau)$ (see text for details).}
 \label{AcousticSHG}
 \end{centering}
\end{figure}

In this study, we consider the influence of the acoustic pulses
generated in the first 16~nm-thin iron layer on the SHG output in
the second one, see Fig.~\ref{AcousticSHG}. The elementary and
oversimplified understanding of photo-acoustic generation suggests
that femtosecond pump pulses are primarily absorbed in iron which
then generates acoustic pulses propagating in both directions. A
3~nm-thin layer of gold aiming to protect the upper iron layer
from oxidation acts as an acoustic delay line and results in the
bi-polar acoustic pulse emitted into the thicker (49~nm) epitaxial
Au(001) layer. An approximate pulse shape is shown in
Fig.~\ref{AcousticSHG}(a) for three distinct time moments
$\tau_1,\tau_2$ and $\tau_3$ corresponding to the arrival times of
the sharp acoustic fronts at the second Fe/Au interface. This
schematic helps to understand the effect of acoustic pulses on the
linear reflectivity and SHG output from the second iron layer with
the thickness $d_{Fe}=15$~nm, see Fig.~\ref{AcousticSHG}(b) and
their high-resolution temporal zoom in Fig.~\ref{AcousticSHG}(c).
A few remarkable observations can be inferred from these
measurements. First, the nonlinear acoustically-induced modulation
$dI_{2\omega}/I_{2\omega}$ appears to be almost an order of
magnitude larger as compared to the linear acoustic modulation
$dR(t)/R$. Second, both optical signals are precisely correlated
with the arrival times of acoustic fronts at the interfaces.

Interestingly, transient $dI_{2\omega}/I_{2\omega}$ (dSHG/SHG in
Fig.~\ref{AcousticSHG}) is a continuous function of time. A
comparison of the SHG output variations with the linear
reflectivity measurements in Fig.~\ref{AcousticSHG}(c) clearly
shows that the maximum compression (dilution) of the second iron
layer correspond to the minimum (maximum) of both SHG intensity
and reflectivity modulation. However, the fact that the kinks in
the transient variations of the SHG signal are significantly
narrower than those in linear reflectivity data demonstrates the
extreme sharpness of the fronts of the acoustic pulses.

In centrosymmetric media SHG may arise due to electro-dipole
(interface) or magneto-dipole and electro-quadrupole (bulk)
contributions. It is commonly believed that the bulk contributions
in metals are small as compared to the interface terms
\cite{RaschkeChapter7,Heinz,WangKauranen}. On the other hand,
inversion symmetry in the bulk can be broken, for example, by a
gradient of strain resulting in the allowed volumetric
electro-dipole contribution. Whereas static strain gradients
emerge from the growth on a lattice-mismatched substrate, dynamic
strain gradients can be induced by ultrashort acoustic pulses.
Another possible mechanism relies upon transient acoustic
modulation of the surface properties of Fe, including the
susceptibility tensor $\chi^{(2)}$. In any case, the total SHG
output is produced by an interference of multiple SHG sources,
whereas the interference conditions are strongly affected by the
propagating acoustic pulses.

Despite these ambiguities, the shape of acoustically-induced SHG
output modulation can be described within a simple model which
considers an interference of the two surface SHG contributions
$E^{2\omega}_1{\rm exp}(i\phi_0)$ and $E^{2\omega}_2$ generated at
the Au/Fe and Fe/MgO interfaces. Here $\phi_0$ denotes the phase
difference between these two interfering terms suggesting that the
total SHG intensity is proportional to
$I_{2\omega}\propto|E^{2\omega}_1{\rm
exp}(i\phi_0)+E^{2\omega}_2|^2$. An acoustic pulse $\eta(z,\tau)$
propagating through the iron layer modulates this optical phase
$\phi(\tau)=\phi_0+\Delta\phi(\tau)$ according to
\begin{equation}
\label{DeltaPhi}
\Delta\phi(\tau)\propto\frac{2\pi}{\lambda}\int_0^{d_{\rm
Fe}}\eta(z,\tau)dz\,,
\end{equation}
where the integration is performed over the entire Fe layer. As
such, we obtain
$dI_{2\omega}=-(2E^{2\omega}_1E^{2\omega}_2\sin\phi_0)\Delta\phi(\tau)$.
Using the acoustic pulse shape shown in Fig.~\ref{AcousticSHG}(a)
and Eq.~(\ref{DeltaPhi}) we have calculated transient phase
variations $\Delta\phi(\tau)$. With an appropriate scaling and
vertical offset, this simulated phase shift demonstrates
surprising similarity to the experimental SHG signal in
Fig.~\ref{AcousticSHG}(c). We note, however, that at present the
underlying origin of this phase modulation remains unclear. Due to
the nonlinear character of the discussed interference, this phase
modulation is acquired by either the driving fundamental fields
$E_1$, $E_2$ or the emitted SHG field at the double frequency
$2\omega$.

From a practical perspective, these preliminary experimental
observations and theoretical modeling hold high potential in
application to the nonlinear plasmonics and magneto-plasmonics.
They can serve as a first step to test the effective interface
approximation for complex multilayer structures, where a simple
interference model can be generalized to include strain-induced
modification of bulk SHG terms dominating in semiconductor and
dielectic media \cite{ZhaoPRB2011,HuberPRB2015}.

A more general phenomenological description using a single
time-dependent nonlinear effective susceptibility $\chi^{(2)}_{\rm
eff}(t)$ for the upper interface in Fig.~7(b)  would be also
useful to understand the physical limits of ultrafast acoustic
modulation in the nonlinear plasmonic switch sketched in Fig.~1.

\section{Fabrication and characterization of multilayer structures}

The quality of the samples determines the effects which can be experimentally observed. For instance, for the plasmonic experiments, the key optimization of the samples consists in the minimization of the sample roughness \cite{McPeakACSPhotonics2015}. In stark contrast, 
ultrafast acoustics turns out to be much more demanding to the
sample quality with many parameters drastically influencing the
observed signal amplitudes and shapes of the transients. Those
parameters include the surface and interface roughness, single- or
polycrystallinity of the samples, grain sizes and grain size
distribution as well as the texture of the grains. For the latter,
due to the large elastic anisotropy of crystalline solids, the
acoustic propagation time in grains with different orientation
would typically differ by 10$\%$ around the average value: The
commonly used value of 3200~m/s for polycrystalline gold differs
from 3450~m/s for (111) and 3150~m/s for (100) crystallographic
directions in gold single crystals.

For the experiments on ultrafast acoustics discussed in this
review, the most appropriate substrates are Al$_2$O$_3$(0001). Due
to their single crystal nature, they assure excellent optical
transparency in the relevant frequency range and offer good
acoustic impedance matching to cobalt. The latter is a good
transducer for ultrafast acoustics because of its high speed of
sound and high melting temperature. Since the details of the layer
stack are already defined for the efficient
magneto-acousto-plasmonics, the remaining optimization of the
sample quality is mainly related to the adjustment of the growth
parameters of the Co layer and Au films, with the target to
improve homogeneity of the respective materials.

One of the key parameters determining the crystalline structure of
Co when grown on Al$_2$O$_3$(0001) is the deposition temperature.
This temperature should be chosen to obtain strong hcp (0001)
texture of Co thin films which should promote the growth of the
(111) textured Au layer. A rather large lattice mismatch of about
14$\%$ between fcc Au(111) and hcp Co(0001) is relieved by
formation of misfit dislocations \cite{LeePRL1989}, which assures
the emergence of laterally extended grains of (111) textured Au on
top of the Co film. We note that the deposition at temperatures of
lower than 200$^{\circ}$C does not result in a clear texture of Co
films on Al$_2$O$_3$(0001) even when rather thick (80~nm) films
are deposited \cite{BrandenburgPRB2009}. At the same time, growth
at temperatures of above 400$^{\circ}$C might favor the Co layer
to arrange in fcc rather than hcp structure, because according to
the equilibrium phase diagram of Co, the hcp to fcc transformation
temperature is 422$^{\circ}$C. In the temperature window between
200$^{\circ}$C and 400$^{\circ}$C, the substrate temperature is
sufficiently large to result in a Co film with hexagonal crystal
structure and a full epitaxial relation
Co(0001)$<10-10>||$Al$_2$O$_3$(0001)$<11-20>$. Although films
grown at higher temperatures typically possess larger grain sizes,
deposition of Au on top of Co at 400$^{\circ}$C might result in
the intermixing at the interface. In accordance with
Ref.\cite{BroederPRL1988,YamaneJMMM1993}, annealing in H$_2$/N$_2$
atmosphere at temperatures up to 300$^{\circ}$C allows to avoid
alloying. However, vacuum annealing of Co/Au multilayers up to
250$^{\circ}$C leads to the partial loss of the structural
coherence which might be indicative for the Co-Au mixture
formation \cite{GubbiottiTSF2003}. Based on these considerations,
the appropriate temperature window for the sample growth is
between 200 and 250$^{\circ}$C.

The deposition of Co and Au thin films is carried out at
250$^{\circ}$C using dc-magnetron sputtering in a high vacuum
chamber with a base pressure of $10^{-7}$~mbar. Double-side
epipolished 500~$\mu$m-thick Al$_2$O$_3$(0001) single crystals
(Crystec GmbH) with lateral dimensions of 10 x 10~mm$^2$ were used
as a substrate. The tilt of the c-axis of the Al$_2$O$_3$(0001)
was less than 0.3$^{\circ}$. The substrates were fixed to the
sample holder using metal clamps assuring good thermal contact.
Before the deposition, the substrates were outgassed for 30 min at
250$^{\circ}$C. Ar was used as a sputter gas (Ar pressure is
3.5x10$^{-3}$~mbar). The deposition rate of Co and Au was set to
0.2~$\AA$/s and 2~$\AA$/s, respectively. The thickness of Co layer
was varied between 2 and 35~nm. The thickness of Au film was
chosen between 120 and 780~nm.

\begin{figure}[h]
  \includegraphics[width=8cm]{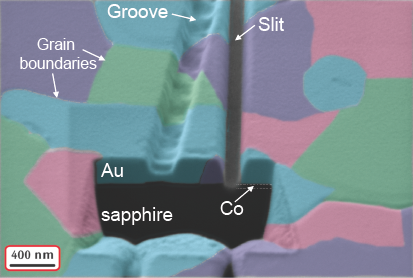}
    \caption{All investigated gold films grown by magnetron sputtering at 250$^{\circ}$C displayed (111) texture and a large grain size of the order of 1~$\mu$m.  Small changes in the SEM images from secondary electrons were color-coded to show different grains.}
 \label{SampleDefects}
\end{figure}

The structural analysis of the samples with Co(2~nm; 4~nm;~6 nm) /
Au(200~nm) has been performed by X-ray diffraction (XRD) in
Bragg-Brentano geometry using Cu K$_{\alpha}$ irradiation
(wavelength: 1.54~$\AA$) on a laboratory diffractometer (XRD 7,
Seifert-FPM). Due to the small thickness, the diffraction peaks of
Co are not resolved. Independent of the thickness of Co, a clear
diffraction peak at 38.35$^{\circ}$  is observed characteristic of
(111) textured Au films. Although the films are strongly (111)
textured, the in-plane orientation of the Au crystallites is
random, as can be seen when analyzing the scanning electron
microscopy (SEM) image (Fig.~\ref{SampleDefects}). Different
in-plane orientation of the crystallites is shown using false
colors. The size of an individual crystallite is in the range of
few hundreds of nanometers. Furthermore, the cross-section of the
film was studied by cutting it using focused-ion beam etching. We
conclude that each grain is uniform in thickness due to the
excellent homogeneity in the SEM contrast. This finding has a
strong implication for the analysis of the propagation of acoustic
pulses in the film: as the gold film is (111) textured, the
propagation velocity of an acoustic pulse along the film thickness
can be taken as constant and equal to the 3450~m/s, the
characteristic value for the a single crystal with (111)
orientation.

\section{Discussion and future research}

In this study we did not discuss any investigations related to the
ultrafast magnetization dynamics \cite{KirilyukReview2010}. The
phenomenon of ultrafast laser-induced demagnetization, either by
the direct optical excitation \cite{Beaurepaire96PRL76} or
mediated by the ultrafast transport of non-equilibrium hot
electrons across the (noble metal)-ferromagnet interface
\cite{MelnikovPRL2011,RudolfNatComm2012,Eschenlohr2013,VodungboSR2016,ShalagatskyiArxiv2015},
results in the abrupt decrease in the length of magnetization
vector. This phenomenon occurs on a sub-picosecond time scale and
can even result in ultrafast magnetization reversal in
ferrimagnets \cite{RaduNat2011,OstlerNC2012}.

Another option of the laser-induced magnetization control consists
in changing the direction of magnetization as evidenced by the
excitation of the ferromagnetic resonance (FMR) precession. FMR
precession \cite{ZhangPRL02,vanKampen2002,Bigot05ChemPhys318} as
well as excitation of spatially inhomogeneous (magnon)
precessional modes in ferromagnetic thin films
\cite{vanKampen2002,ShalagatskyiArxiv2015} originate from the
ultrafast thermally-induced changes in the magneto-crystalline
anisotropy.

Very recently, it became typical to induce FMR precession via
magneto-elastic interactions in ferromagnets, when excited by
picosecond acoustic strain pulses
\cite{Scherbakov10PRL105,Thevenard10PRB82,Kim12PRL109}. So far the
acoustically induced FMR precession in thin ferromagnetic films
was observed only when the symmetry was broken by the non-zero
out-of-plane component of the external magnetic field
\cite{Scherbakov10PRL105,Thevenard10PRB82,Kim12PRL109}. In the
analogous experiments with ultrafast surface acoustic waves (SAW,
excited by fs-laser pulses in transient grating geometry
\cite{NelsonJAP1982}), the most efficient excitation of FMR
precession was achieved by tilting the external magnetic field
with respect to the SAW wave vector
\cite{Janusonis2015,Janusonis2016}. Further, a laser excitation of
a coupled magneto-elastic mode with an extreme lifetime of more
than 25 ns and strongly nonlinear behaviour was demonstrated in
ferromagnetic dielectric FeBO$_3$ \cite{AfanasievPRL14}.

Purely acoustic magnetization switching was predicted
theoretically for single-crystal thin film of magnetostrictive
metal Terfenol-D excited with picosecond acoustic pulses
\cite{KovalenkoPRL2013} or semiconductor (Ga,Mn)(As,P) driven by
quasi-monochromatic SAW transients \cite{ThevenardPRB2013}.
Experimentally, acoustic magnetization switching has only been
demonstrated in ferromagnetic nanoparticles on the nanosecond time
scale \cite{SampathArxiv2016}. From all these observations we
conclude that in the nonlinear plasmonic switch sketched in
Fig.~1, notably with an in-plane magnetic field applied, the
magneto-acoustic effects are unlikely to play a major role.

The substantial modulation of the nonlinear optical properties of
1.8\% can be further enhanced by functionalizing the structure
with a monolayer of semiconductor quantum dots (QDs). QDs are
particularly promising for nonlinear acousto-magneto-plasmonics as
they feature configurable "atom-like" optical excitation spectrum
because they offer the ability of engineering of the electron and
hole wave functions via the QD size $R_{\rm QD}$.
Nonlinear initialization and readout of singly-charged QDs can be
harvested for ultrafast quantum optical functionality
\cite{Sotier09}. It is known that a linear as well as a nonlinear
response in these systems, typically analyzed in the dipole
approximation which exploits the approximation $R_{\rm
QD}\ll\lambda$, is intimately linked to the details of the spin
configuration of the few-electron states as well as the crystal
structure and the geometry of the nano-emitter
\cite{Efros96,Huneke11}. Application of controlled acoustic
transients with considerable enhancement in the elastic field
gradients on the length scales comparable to $R_{\rm QD}$ would
thus allow for controlled studies of the response of the quantum
system beyond the dipole approximation \cite{ZuritaJOSA02}. In
other words, if the acoustic pulse can significantly perturb the
electron eigenstates in the quantum dot, new fascinating
electromagnetic effects of substantially non-local nature can be
expected.
This nonlinear coupled regime has not been well explored so far
and might provide new physical insights and therefore inspire
novel photonic applications.
As a parting example we speculate that by subjecting colloidal
quantum dots \cite{Roo14} to intense acoustic transients such as
those reviewed here, one might be able to drive structural phase
transition \cite{Alivisatos96} in these crystal on ultrafast
timescales. As a result, one would be able to directly switch and
control the brightness of the quantum emitter \cite{Efros96} on
femtosecond time scales.

Beyond the SHG response, higher nonlinearities such as third-order
susceptibility-driven magnetic third harmonic generation (mTHG)
can be exploited to great effect. Indeed, mTHG allows to evaluate
bulk (as compared to surface mSHG) nonlinearities and also draw
analogies to the giant magnetoresistance properties in
ferromagnetic nanostructures \cite{AktsipetrovPRB2006}. In our
case, for the THG frequency to fall in the visible spectral range,
one would need to apply pump excitation with a wavelength in the
near-infrared spectral range \cite{HankePRL09,HankeNanoLett12}.
Employing the telecom wavelengths ($\lambda=$ 1.3 or 1.5~$\mu$m)
would allow to simultaneously detect mSHG and mTHG in the visible
spectral range, while exploiting mature technology of
ultrabroadband femtosecond fiber lasers \cite{Brida2014}.
Moreover, in this case SPP excitation at the first and the second
harmonic frequencies will occur at nearly the same angle and as a
result would lead to a more favorable phase-matching condition
between the SPPs, $k^{2\omega}_{\rm spp}\simeq 2k^{\omega}_{\rm
spp}$. Moving to even longer wavelengths and using intense pulses
of THz radiation as an effective stimulus may enable efficient
high-order harmonic generation with SPPs.

To summarize, in this paper we have discussed some recent
developments in the field of the nonlinear
acousto-magneto-plasmonics, which are necessary for understanding
of the fundamental interactions and the associated time and length
scales of the nonlinear plasmonic switch sketched in Fig.~1. The
extension of the concepts presented here to hybrid plasmonic
structures as well as to semiconductor nanostructures, together
with shifting the experimental focus to lower photon energies and
higher-order optical nonlinearities hold high potential for future
research.



\subsection{Acknowledgments}
Authors thank A. Alekhin (FHI Berlin) for performing the nonlinear
SHG-measurements shown in Fig.~11 and T. Thomay (SUNY at Buffalo)
for focussed ion beam milling of the plasmonic
microinterferometers and taking a false-color SEM image of
Au/Co/sapphire structure in Fig.~12. They also acknowledge
stimulating discussions with U.~Woggon, A.~Leitenstorfer, A.
Kirilyuk, Th. Rasing, T.V. Murzina, R.~Bratschitsch, R.~Tobey, A.
Garcia-Martin, V.S.~Vlasov, A.M.~Lomonosov, V. Juv\'{e},
P.~Ruello, V.E.~Gusev, I.~Radu, U.~Bovensiepen, T.~Kampfrath,
M.~Bargheer, P. Gaal and M.~Wolf.

Funding from {\it Nouvelle \'{e}quipe, nouvelle th\'{e}matique}
"Ultrafast acoustics in hybrid magnetic nanostructures" and {\it
Strat\'{e}gie internationale} "NNN-Telecom" de la R\'{e}gion Pays
de La Loire, DFG (TE770/1), ANR-DFG "PPMI-NANO" (ANR-15-CE24-0032
\& DFG SE2443/2), ANR "UltramoX" (ANR-14-CE26-0008), U.S. DOE
Grant no. DE-FG02-00ER15087, U.S. NSF Grant no. CHE-1111557, the
European Research Council (FP7/2007-2013) / ERC grant agreement
no. 306277 and {\it Alexander von Humboldt Stiftung} is highly
appreciated.

\section{References}
\bibliography{JOP_Bibliography}

\end{document}